\documentclass[english,prd,superscriptaddress,nofootinbib,
               preprintnumbers,twocolumn,noshowpacs]{revtex4}
\usepackage{amsmath,amssymb,amsthm,mathrsfs,graphicx,bm,color} 

\newcommand{\be}{\begin{equation}}
\newcommand{\ee}{\end{equation}}
\newcommand{\bea}{\begin{eqnarray}}
\newcommand{\eea}{\end{eqnarray}}

\renewcommand{\d}{{\rm d}}

\newcommand{\fnl}{f_{\rm{NL}}}

\newcommand{\mpl}{M_{\rm pl}}
\newcommand{\s}{\sigma}
\newcommand{\p}{\phi}
\renewcommand{\P}{{\mathcal P}}
\renewcommand{\O}{{\mathcal O}}

\newcommand{\C}{{\mathcal C}}
\newcommand{\Q}{{\mathcal Q}}

\newcommand{\F}{{\mathcal F}}

\newcommand{\ep}{\epsilon}
\newcommand{\gp}{\gamma_{\p}}

\newcommand{\gpb}{{\gamma_\p^{\rm B}}}

\newcommand{\gpc}{{\gamma_\p^{\rm C}}}
\newcommand{\gpd}{{\gamma_\p^{\rm D}}}
\newcommand{\gsc}{{\gamma_\s^{\rm C}}}
\newcommand{\gsd}{{\gamma_\s^{\rm D}}}

\begin{document}
\title{General analytic predictions of two-field inflation and perturbative reheating}

\author{Joseph Elliston}
\email{j.elliston@sussex.ac.uk}
\affiliation{Astronomy Centre, University of Sussex, Falmer, Brighton BN1 9QH, UK.}

\author{Stefano Orani}
\email{stefano.orani@unibas.ch}
\affiliation{Department of Physics, University of Basel, Klingelbergstrasse 82, CH-4056 Basel, Switzerland.}

\author{David J. Mulryne}
\email{d.mulryne@qmul.ac.uk}
\affiliation{School of Physics and Astronomy, Queen Mary University of London,\\
Mile End Road, London, E1 4NS,  UK.}

\begin{abstract}
The observational signatures of multi-field inflation will generally evolve as the Universe reheats. We introduce a general analytic formalism for tracking this evolution through perturbative reheating, applicable to two field models with arbitrary separable potentials. The various transitions, including the onset of scalar field oscillations and the reheating of each field, can happen in different orders and on arbitrary hypersurfaces. The effective equations of state of the oscillating fields are also arbitrary. Nevertheless, our results are surprisingly simple. Our formalism encapsulates and generalises a huge range of previous calculations including two-field inflation, spectator models, the inhomogeneous end of inflation scenario and numerous generalised curvaton scenarios.
\end{abstract}

\maketitle

\section{Introduction}

The presence of isocurvature modes in the early Universe has a profound influence on the evolution of cosmic observables. During the slow-roll inflationary phase, this effect has been tracked analytically by Vernizzi and Wands~\cite{Vernizzi:2006ve}, extending earlier work~\cite{GarciaBellido:1995qq}, for two-field models with arbitrary sum-separable potentials. Their results apply to models where reheating occurs suddenly, such as two-field Hybrid inflation where inflation is terminated by the destabilisation of a third heavy field~\cite{Alabidi:2006wa, Alabidi:2006hg, Sasaki:2008uc, Byrnes:2008zy}. In such cases, the isocurvature modes are assumed to decay almost instantaneously due to the onset of a phase of thermal equilibrium. It is possible that this process of sudden reheating may occur on a non-uniform density hypersurface. This is known as the Inhomogeneous End of Inflation scenario~\cite{Bernardeau:2002jf,Lyth:2005qk,Salem:2005nd,Alabidi:2006wa,Bernardeau:2007xi,Sasaki:2008uc,Naruko:2008sq,Huang:2009vk} and has been recently summarised and constrained by Planck data in ref.~\cite{Elliston:2013afa}. Since the method of Vernizzi and Wands~\cite{Vernizzi:2006ve} tracks perturbations until the end of inflation, their work is equally applicable to any two-field separable potential where observable statistics attain constant conserved values before the end of inflation. This happens when the isocurvature modes decay during inflation and the system reaches an `adiabatic limit'~\cite{Meyers:2010rg,Elliston:2011dr}. 

More generally, isocurvature modes may be present after inflation, and inflation may not end suddenly. This will mean that both the inflationary and the post-inflationary phases will cause the cosmic observables to evolve. As emphasised in Elliston et al.~\cite{Elliston:2011dr}, robust predictions can only then be generated if the primordial perturbations are tracked through the post-inflationary phases, either until isocurvature modes do finally decay, or until the time at which the system is observed. In particular, it is of vital importance to consider how cosmic observables may be sensitive to the process by which the Universe reheats. In addition to the sudden non-perturbative reheating mechanism present in Hybrid inflation, perturbative reheating can also be considered. In this case an oscillating field dissipates its energy more gradually, often significantly after the end of inflation. Therefore any analytic calculation for a model exhibiting this behaviour must give due consideration to the post-inflationary phases. The most prominent example of the effect of perturbative reheating on the generation of primordial perturbations is the `curvaton' scenario~\cite{Lyth:2001nq,Moroi:2001ct,Enqvist:2001zp,Sasaki:2006kq}. 

A third method to reheat the Universe is preheating, driven by explosive resonant particle production. Like the end of Hybrid inflation, this form of reheating is sudden and non-perturbative. If the field into which energy is being transferred is light, it can have a significant effect on the evolution of perturbations~\cite{Bassett:1999cg,Finelli:2000ya,Chambers:2007se, Bond:2009xx}. Since such effects must be studied using lattice simulations, however, we shall not consider this behaviour in our analytic work, restricting our attention instead to perturbative reheating.

The goal of this paper is to augment the work of Vernizzi and Wands~\cite{Vernizzi:2006ve} to derive a simple and usable set of analytic formulae that allow us to understand how perturbative reheating modifies cosmic observables. Our method can be applied to any two-field sum-separable model. We draw upon and generalise many different calculations that have been tailored to more specific scenarios; these scenarios may then be recovered as particular limits of our work. In particular, we account for the Modulated Reheating scenario~\cite{Dvali:2003em,Zaldarriaga:2003my,Kofman:2003nx,Vernizzi:2003vs,Bernardeau:2004zz,Ichikawa:2008ne,Suyama:2007bg} where the hypersurface on which a particular field reheats is directly dependent on the value of an additional light field, and for a multitude of possible generalisations to the standard curvaton scenario. These include modulation of the onset of curvaton oscillations~\cite{Kawasaki:2011pd, Kawasaki:2012gg}, modulation of the reheating hypersurface for the curvaton~\cite{Enomoto:2012uy,Assadullahi:2013ey,Langlois:2013dh,Assadullahi:2012yi,Enomoto:2013qf}, and modulation of the reheating hypersurface for the inflaton~\cite{Choi:2012te}. See refs.~\cite{Elliston:2013afa,Vernizzi:2003vs,Alabidi:2010ba} for other works drawing links between these scenarios.

Our work follows a considerable amount of recent effort in this field. Numerical studies~\cite{Elliston:2011dr,Leung:2012ve} have demonstrated that altering the decay widths of one or both fields causes the local shape bispectrum parameter $\fnl$ to evolve. Subsequently, Meyers and Tarrant~\cite{Meyers:2013gua} demonstrated that this evolution was caused by the same physical mechanism operating behind the curvaton scenario; the curvature perturbation is modified by the relative redshifting of two components with different equations of state. They showed this by providing general analytic formulae that describe the evolution of perturbations during the perturbative reheating phase. The effect of perturbative reheating has also been discussed in the context of particular models in refs.~\cite{Enomoto:2013bga,Huston:2013kgl}.  

The formalism that we develop in this paper is more general than that of Meyers and Tarrant~\cite{Meyers:2013gua}. We allow for reheating to occur on arbitrary hypersurfaces, allowing us to consider the modulated reheating scenarios discussed above. Since there is no agreed analytic method for fixing the hypersurfaces of oscillation onset, we also leave these hypersurfaces arbitrary. This ensures that our formalism, instead of being limited by the lack of a well-defined oscillation criterion, may be used as a testing-ground for choosing between methods of defining the oscillation hypersurfaces. The explicit nature of our formalism also makes it very simple to interpret and to obtain standard limiting cases.

We split the evolution into four phases, labelled A, B, C and D, which are summarised in figure \ref{fig:summary}. At any particular time, the two components involved will be described either as scalar fields undergoing generalised slow-roll, or as fluids with a constant equation of state. Phase A contains two slow-roll fields and is therefore a reiteration of the results of ref.~\cite{Vernizzi:2006ve}. The four-phase picture we employ is a choice which covers the vast majority of models encountered in the literature, but it is not exhaustive. However, the basic tools that we develop can be assembled by the user to tackle any similar problem even if it falls outside of our picture. \\

\noindent {\bf Paper structure}: In \S\ref{sec:background_theory} we introduce the necessary background theory. \S\ref{sec:deltaN} then shows how the curvature perturbation $\zeta$ may be computed in terms of horizon exit field perturbations and provides the principle result of this paper. \S\ref{sec:ns} then provides a general discussion regarding predictions for the scalar spectral index and the tensor to scalar ratio. \S\ref{sec:second_order} provides the ingredients for deriving predictions at second order, which may be assembled to suit specific problems as we show in \S\ref{sec:models}. We conclude in \S\ref{sec:conclusions} and some supporting formulae are given in the appendix.

\section{Background theory}
\label{sec:background_theory}

\begin{figure*}
\centering
\includegraphics[width=1.5\columnwidth]{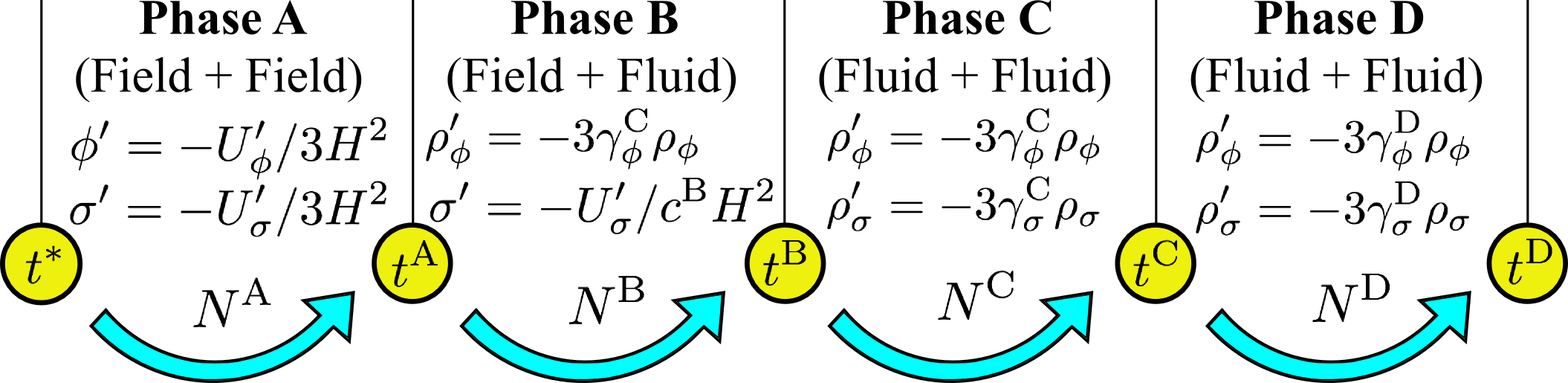}
\caption{Summary of the four stages of our calculation. Phase A persists for $N^{\rm A}$ efolds between the hypersurfaces at times $t^*$ and $t^{\rm A}$. Note that since $\gp$ does not change at $t^{\rm B}$ then $\gpb = \gpc$ and we use the latter throughout.}
\label{fig:summary}
\end{figure*}

\noindent {\bf Observable quantities}: 
The primordial curvature perturbation on uniform density spatial hypersurfaces is denoted by $\zeta$ (e.g.~\cite{Malik:2008im}). The statistical properties of $\zeta$ are constrained by observations and are the basis for the observational quantities we are interested in. They are commonly measured in terms of the power spectrum and bispectrum (and trispectrum), which are defined as 
\begin{eqnarray}
\label{spectra}
\langle\zeta_{\mathbf k_1}\zeta_{\mathbf k_2}\rangle &\equiv& (2\pi)^3
\delta^3 ({\mathbf k_1} + {\mathbf k_2}) P_{\zeta}(k) \, , \nonumber \\
\langle\zeta_{\mathbf k_1}\,\zeta_{\mathbf k_2}\,
\zeta_{\mathbf k_3}\rangle &\equiv& (2\pi)^3 \delta^3 ( {\mathbf k_1}+{\mathbf k_2}+
{\mathbf k_3}) {\cal B}_\zeta( k_1,k_2,k_3)  \,, \nonumber
\end{eqnarray}
respectively, where for the two point statistics we define the common magnitude $k = |\mathbf k_1| = |\mathbf k_2|$. The power spectrum is commonly written in the dimensionless form $P_\zeta(k) = (2\pi^2 / {k}^3) {\cal P}_{\zeta}(k)$, which has mild scale-dependence parametrized by the scalar spectral index $n_\zeta -1= \d \ln {\cal P_\zeta}/ \d \ln k$. For the canonical models of interest in this paper, only the local shape of non-Gaussianity is relevant and this can be written in terms of the weakly scale-dependent $\fnl$ parameter as
\bea
\label{fnl}
{\cal B}_{\zeta}(k_1,k_2,k_3)&=&\frac{6}{5} \fnl \big[ 
P_\zeta(k_1) P_\zeta(k_2) 
+ \mathrm{2~perms} \big].
\eea

\noindent {\bf The $\delta N$ formalism:} In order to follow the evolution of $\zeta$ on super-horizon scales, and calculate its statistics, we employ the separate universe approach to perturbation theory~\cite{Lyth:1984gv,Wands:2000dp}, and the $\delta N$ 
formalism~\cite{Starobinsky:1986fxa,Sasaki:1995aw,Lyth:2005fi}. This equates $\zeta$ to the variation in the number of efolds $N$ between a flat hypersurface at horizon exit, which we label `$*$', and a subsequent uniform density hypersurface. For a system of scalar fields $\varphi_i$ we can then expand $\zeta$ in terms of field perturbations at horizon exit as
\be
 \label{eq:deltaN}
\zeta \equiv \delta N = N_{,i} \delta \varphi_i^*
+ \frac{1}{2}  N_{,ij} \delta \varphi_i^* \delta \varphi_j^* +  \dots\,,
\ee
where we employ the summation convention and $N_{,i} = \partial N / \partial \p_i^*$. The $\delta N$ formalism allows us to write simple expressions for the cosmological parameters $\P_\zeta$, $n_\zeta$, $\tilde r$ and $\fnl$, where $\tilde r$ is the ratio of the tensor and scalar power spectra. Here, since we only consider two fields $\phi$ and $\s$, it is helpful to define the parameter $R=N_{,\s}^2/N_{,\phi}^2$ such that $R \ll 1$ means that $\phi$ dominates the linear order statistics and $R \gg 1$ implies the $\s$ field dominates. For the case of sum-separable potentials one then finds~\cite{Vernizzi:2006ve}
\begin{subequations}
\begin{align}
\P_\zeta &= \, N_{,\p}^2 \big(1 + R\big)  \P_{\delta \phi} \,, 
\label{eq:power} \\
\tilde r &= \frac{8}{\mpl^2 N_{,\p}^2 (1+R)} \,, \label{eq:r} \\
n_\zeta -1 &= -2 \ep^* + 2 \, \frac{R \, \eta_{\s \s}^*  + \eta_{\p \p}^* - \mpl^{-2} N_{,\p}^{-2}}{1+R} ,
\label{eq:ns} \\
\!\!\!\!\! \frac{6}{5} \fnl &= \frac{1}{(1+R)^2} \bigg(
\frac{N_{,\p \p}}{N_{,\p}^2} + 2 \frac{N_{,\p \s}}{N_{,\p}N_{,\s}}  R + \frac{N_{,\s \s}}{N_{,\s}^2} R^2 \bigg)
\label{eq:fnl} , \!\!
\end{align}
\end{subequations}
where $\P_{\delta \p} = \langle \delta \p^* \,\delta \p^* \rangle = 
\langle \delta \s^* \,\delta \s^* \rangle = H_*^2 / 4\pi^2$. \\

\noindent {\bf Scalar field evolution}: We prescribe that the fields $\p$ and $\s$ are not directly coupled. This requires that the inflationary potential has a sum-separable form $V(\p,\s) = U_\p(\p) + U_\s(\s)$. The scalar fields then each evolve as
\begin{equation}
\label{eq:full_evo}
	\ddot \p + 3 H \dot \p + U_\p' = 0 \,,
\end{equation}
where an overdot denotes differentiation with respect to coordinate time. Note that we use primes in two distinct ways: when applied to a potential $U$ they denote a partial derivatives $U_\p' = \partial U_\p / \partial \p$, and for all other quantities they denote derivatives with respect to the number of efolds $N$ as $\s' = \d \s / \d N$. Kawasaki et al.~\cite{Kawasaki:2012gg} showed that the dynamics associated with eq.~\eqref{eq:full_evo} approach those of the linear-order attractor solution
\begin{equation}
	c H \dot \p = -U_\p' \,,
	\label{eq:attractor}
\end{equation}
where $c = 3+\ep $ is a function of the background cosmology. One therefore finds values $c=3$ for de Sitter, $c=5$ for radiation and $c=9/2$ for matter background equation of state. Defining the generalised slow-roll parameters
\begin{equation}
\ep_\p = \frac{\mpl^2 {U_\p'}^2}{2 \rho^2}\,, \qquad
\eta_{\p\p} = \frac{\mpl^2 {U_\p''}}{\rho} \,,
\end{equation}
where $\rho$ is the total density, one requires that $|\eta_{\p\p}| \ll c/3$ for the solution~\eqref{eq:attractor} to be valid.\\

\noindent {\bf Oscillating fields}: 
Once a scalar field nears the minimum of its potential then the attractor solution \eqref{eq:attractor} breaks down and the field begins oscillations. If the minimum of the $U_\p$ potential can be described perturbatively by the monomial $U_\p = \lambda \p^p$, where $p$ is a constant, then the time-averaged behaviour of the oscillating $\p$ field behaves as a fluid $\rho_\p$ which redshifts according to the equation $\rho_\p' = -3 \gamma_\p \rho_\p$ where $\gamma_\p$ is the equation of state which we presume to be constant. For a quadratic minimum with $p=2$ the fluid evolves like dust with $\gamma_\p = 1$, whereas for a quartic minimum with $p=4$ the fluid evolves like radiation with $\gamma_\p = 4/3$.\\

In this phase we can model the field as a fluid, which significantly simplifies analytic calculations. This fluid approximation is also useful for numerical work, since it avoids the computational expense of tracking an oscillating field over many efolds. We include this in our numerical implementation for scenarios involving very many oscillations by allowing such an oscillating field to decay into an effective fluid with the same equation of state. The equations which allow this decay to proceed, while conserving energy, are 
\begin{subequations}
\label{eq:effective_decay}
\begin{align}
\ddot \p  + 3 H \dot \p  + \Gamma^{\rm eff}_\phi \dot \p + U_\p' &= 0 \,, \\
\dot \rho_\p + 3 \gamma_\p H \rho_\p - \Gamma^{\rm eff}_\p \dot \phi^2 &= 0 \,,
\end{align}
\end{subequations}
where the $\Gamma^{\rm eff}_\p$ parameter controls when the oscillating field decays. This follows because most of the energy in the oscillating field is transferred to the effective fluid when $\Gamma^{\rm eff}_\phi \sim H$, and so we must select a value of $\Gamma^{\rm eff}_\phi$ to ensure that this transfer happens deep into the oscillating regime. Whilst it is not technically correct to include this term in the equations of motion before the oscillations begin~\cite{Leung:2012ve}, our method has the advantage of avoiding spurious issues associated with choosing a time to turn this term on. Any inaccuracy incurred can be damped exponentially by decreasing $\Gamma^{\rm eff}_\p$, which allows more oscillations before the decay. Where we include these effective fluids in our numerical implementation, we ensure accuracy by verifying that $\Gamma^{\rm eff}_\p$ is sufficiently small that our results are insensitive to the precise value. The $\s$ field can oscillate as an effective fluid in an analogous way. \\

\noindent {\bf Perturbative reheating}: 
Perturbative reheating is a phase of evolution in which an oscillating scalar field gradually loses energy to other particles~\cite{Linde:2005ht}, and eventually, perhaps through further decays, to the particles of the Standard Model. From the cosmological perspective, perturbative reheating can be described via effective field equations that couple the fields to their decay products.  

The simplest case has each field coupled to a single decay product which can be modelled as a fluid with a fixed equation of state. For the $\p$ field, this coupling is mediated through a parameter $\Gamma_\p$, and similarly for $\s$. Note that $\Gamma_\p$ is a physical parameter that alters the predictions of the model, unlike $\Gamma^{\rm eff}_\phi$ which does not. Given that the decay occurs when the oscillating field is acting like an effective fluid $\rho_\p$, an equivalent description is the decay from one fluid (the effective oscillating field) to another fluid. We assume that the decay product from one field does not interact with the decay product of the other field. 

Reheating actually takes place over some time interval in which both the effective fluid $\rho_\p$ and its decay product must be modelled. This is captured by simple numerical implementations, but our analytic calculation employs the assumption of an instantaneous transition. This is a good approximation since the majority of the density in the oscillating field is always converted to the decay fluid when $\Gamma_\phi \sim H$~\cite{Malik:2006pm}. Thus the reheating of $\p$ can be analytically modelled by a single fluid $\rho_\p$  that undergoes an instantaneous change in its equation of state $\gamma_\p$ on the reheating hypersurface.  If $\Gamma_\p$ is truly a constant, the reheating surface will be a uniform density hypersurface, but in practice $\Gamma$ might be modulated by one of the fields leading to a modulated reheating scenario. Our analytic methods fully account for this possibility.\\

\noindent {\bf Four-phase setup}: 
We calculate $\delta N$ for an evolution consisting of four phases as summarised in figure~\ref{fig:summary}. Without loss of generality we presume that the $\p$ field is the first to begin oscillations at a time $t^{\rm A}$, whilst the $\s$ field begins to oscillate at a time $t^{\rm B}$. At a later time $t^{\rm C}$ one of these fluids will decay to radiation and the final field decays to radiation at a time $t^{\rm D}$. We do not require that these decays occur in a particular order, but we do presume that such decays are instantaneous and that they only occur after both fields have begun oscillating. This four-phase calculation is a choice that allows us to describe many different models, but it is not exhaustive. However, we stress that predictions can be made for models that do not fit into our four-phase picture, simply by assembling our formalism in a different order. \\

\section{$\bm{\delta N}$ for two fields and fluids}
\label{sec:deltaN}

In this section we derive simple $\delta N$ formulae that apply to all two-component scenarios where the components are not directly coupled. To maintain generality we allow the important transitions (where the fields begin oscillating or the fluids decay to radiation) to occur on arbitrary hypersurfaces. We begin by writing an expression for the total value of $N$ between $t^*$ and $t^{\rm D}$. Differentiating this then provides us with $N_{,\p}$ and $N_{,\s}$ which may then be differentiated to find the higher order derivatives such as $N_{,\s\s}$. Therefore, our linear result for $\delta N$ that traverses both the field and fluid regimes is the principle goal of this paper.\\

\subsection{\bf Deriving an expression for $\bm{\delta N}$. }
Between $t^{\rm A}$ and $t^{\rm D}$ the exact integrability of the $\rho_\p$ equation of motion ensures that we are able to use it as a clock. We find the number of efolds in these phases by integrating the $\rho_\p$ equation of motion to yield
\begin{equation}
N^{\rm B} = \frac{1}{3 \gpc} \ln \frac{\rho_\p^{\rm A}}{\rho_\p^{\rm B}},~
N^{\rm C} = \frac{1}{3 \gpc} \ln \frac{\rho_\p^{\rm B}}{\rho_\p^{\rm C}},~
N^{\rm D} = \frac{1}{3 \gpd} \ln \frac{\rho_\p^{\rm C}}{\rho_\p^{\rm D}}.
\end{equation}
The contribution $N^{\rm A}$ is found by integrating the equation of motion~\eqref{eq:attractor} for the $\p$ field with $c^{\rm A}=3$ due to the scalar field domination. One finds
\begin{equation}
\label{eq:N_stage_1}
N^{\rm A} = 
-\int_*^{\rm A} \! \frac{U_\p}{ \mpl^2 U_\p'} \, \d \p  
-\int_*^{\rm A} \! \frac{U_\s}{\mpl^2 U_\s'} \, \d \s \,,
\end{equation}
where we have used the slow-roll assumption to equate the total density to the summed potential energy as $\rho = U_\p + U_\s$ and we have applied eq.~\eqref{eq:attractor} which relates the evolution of the two fields as $\d \p / U_\p' = \d \s / U_\s'$. 
Combining these results we find the total number of efolds $N = N^{\rm A}+N^{\rm B}+N^{\rm C}+N^{\rm D}$ as
\begin{align}
\label{eq:N_total}
N &= 
-\int_*^{\rm A} \! \frac{U_\p}{\mpl^2 U_\p'} \, \d \p   \,
-\int_*^{\rm A} \! \frac{U_\s}{\mpl^2 U_\s'} \, \d \s  \nonumber \\
& \qquad -\frac{1}{3 \gpc} \ln \frac{\rho_\p^{\rm C}}{\rho_\p^{\rm A}}
-\frac{1}{3 \gpd} \ln \frac{\rho_\p^{\rm D}}{\rho_\p^{\rm C}} \,.
\end{align}
Varying this result provides $\delta N$. The lower boundary terms from the two integrals in eq.~\eqref{eq:N_total} yield the two contributions to $\zeta$ that define the `Horizon Crossing Approximation' (HCA)~\cite{Kim:2006te,Kim:2010ud,Elliston:2011dr,Elliston:2011et} and so we write these terms as 
\begin{equation}
\label{eq:HCA}
\delta N_{\rm HCA} = \frac{U_\p}{\mpl^2 U_\p'}\bigg|_* \delta \p^*  
+ \frac{U_\s}{\mpl^2 U_\s'}\bigg|_* \delta \s^* \,.
\end{equation}
The complete result for $\delta N$ includes terms proportional to perturbations on hypersurfaces $t^{\rm A}$, $t^{\rm B}$, $t^{\rm C}$ and $t^{\rm D}$ as
\begin{align}
\label{eq:dN_1}
\!\!\!\! \delta N &= 
\delta N_{\rm HCA} -\frac{U_\p}{\mpl^2 U_\p'}\bigg|_{\rm A} \delta \p^{\rm A}   
- \frac{U_\s}{\mpl^2 U_\s'}\bigg|_{\rm A} \delta \s^{\rm A} 
\nonumber \\
& +\frac{1}{3 \gpc} \delta \ln \rho_\p^{\rm A} 
+\frac{(\gpc-\gpd)}{3 \gpc \gpd} \delta \ln \rho_\p^{\rm C} 
-\frac{1}{3 \gpd} \delta \ln \rho_\p^{\rm D} \,.
\end{align}
The HCA exploits the simplifying assumption, when it is valid, that all of the perturbations on these later hypersurfaces are negligible and so $\delta N = \delta N_{\rm HCA}$. For this to be the case, isocurvature perturbations must be attenuated during phase A so that the only perturbation at times $t \geq t^{\rm A}$ is a single adiabatic mode. Since an adiabatic perturbation intercepts a foliating hypersurface at a unique phase space point, any model that achieves this adiabatic condition by or before $t^{\rm A}$ will have negligible field perturbations at later times as measured on any such hypersurface.

Eq.~\eqref{eq:dN_1} is complete and technically contains both $N_{,\p}$ and $N_{,\s}$ but before these can be read off we need to relate the perturbations at later times to their values at horizon exit. This is the challenging part of this calculation which occupies the rest of this section, where we relate all perturbations to those at horizon exit in a very general way that does not require any specific choices for the hypersurfaces of the transitions at the times $t^{\rm A}$, $t^{\rm B}$, $t^{\rm C}$ or $t^{\rm D}$, or any specific equations of state for the oscillating fields. These conditions may then be fitted to the problem at hand. We note that we therefore provide all of the tools necessary for investigating the effect of modulated oscillations or modulated reheating in one unified analytic framework. 

Finally, we note that if one wishes to identify $\zeta=\delta N$ then one needs to take the limit where the hypersurface at $t^{\rm D}$ is one of uniform density. We do not automatically make this assertion in our method, partially because it is just as easy to remain general, but also because it is conceivable that one may wish to consider a more general model which includes subsequent phases after $t^{\rm D}$ before the calculation is complete, and these may not match onto the end of phase D on a hypersurface of uniform density. 

\subsection{\bf Accounting for arbitrary hypersurfaces. }

Before we can evaluate $\delta N$ in terms of horizon exit perturbations we must first introduce the technology that will allow us to parametrize the hypersurfaces of oscillation and reheating in full generality. This is simply done by defining four different functions $f^{\rm A}$, $f^{\rm B}$, $f^{\rm C}$ and $f^{\rm D}$, one for each transition, where each of these $f$-functions are dependent on the densities $\rho_\p$ and $\rho_\s$ at their respective times of evaluation and all four $f$-functions equal constants. These then allow us to relate the two perturbations on a given hypersurface, such as
\begin{equation}
\label{eq:arbitrary_gauge}
f_{,\s}^{\rm A}\delta \rho_\s^{\rm A}  + f_{,\p}^{\rm A} \delta \rho_\p^{\rm A} = 0\,,
\end{equation}
where $f_{,\p} = \partial f / \partial \rho_\p$. These terms only ever appear in the ratio $f_{,\p} / f_{,\s}$ and so there is effectively only one parameter here. Note that if one or both of the components are fields then we identify $\rho_\alpha \to U_\alpha$. We also note that in the case of a uniform density hypersurface one finds $f_{,\p} = f_{,\s} = 1$, or a uniform-$\s$ hypersurface has $f_{,\p} = 0$. 

Such relations are useful because they allow us to rewrite all instances of $\s$ perturbations on the given hypersurfaces in terms of $\p$ perturbations, which greatly simplifies the algebra when computing $\delta N$. Another algebraic simplification is made by defining a parameter `$r$' associated to each hypersurface. Physically $r$ is the proportion by which the time-evolution of the hypersurface function $f$ is sourced by the evolution of the $\s$-component, meaning that it has a general form as
\begin{equation}
\label{eq:r_fun}
r = \frac{f_{,\s} \rho_\s'}{f_{,\s} \rho_\s' + f_{,\p} \rho_\p'} \,.
\end{equation}
We note that in general $r$ may take any positive or negative real value, depending on the hypersurface function $f$. The constraints imposed on $f$ in the uniform density case equate to placing bounds on $r$ as $0\leq r \leq 1$. To be concrete, $r$ is calculated fractionally before the hypersurface in question such that there is no ambiguity as to which formula should be used for the evolution of the various densities appearing in eq.~\eqref{eq:r_fun}. We then find the reheating hypersurface at $t^{\rm D}$ has $r^{\rm D}$ of the form
\begin{equation}
\label{eq:rD}
r^{\rm D} = \frac{f_{,\s} \gamma_\s \rho_\s}{f_{,\s} \gamma_\s\rho_\s + f_{,\p} \gamma_\p\rho_\p} \bigg|_{\rm D} \,,
\end{equation}
and $r^{\rm C}$ is identical after relabeling ${\rm D} \to {\rm C}$. We note that $r^{\rm D}$ reduces to the Lyth `$r$' parameter in the standard quadratic curvaton scenario when $\gsd = 1$ and $\gpd = 4/3$ and the hypersurface at $t^{\rm D}$ is uniform density. \\

\noindent{\bf Hypersurface of oscillation onset}:
There is no clear analytic prescription for determining when a field begins to oscillate. In this paper, we shall remain agnostic about this choice and allow oscillations to begin on arbitrary hypersurfaces. One way to achieve this is via the same methodology used for the reheating hypersurfaces. But we now show that it is also useful to define `oscillation factors' $x_\p$ and $x_\s$. These are simply another way to encode the values of the parameters $f_{,\p}$ and $f_{,\s}$. 

Considering the hypersurface at $t^{\rm A}$ where the $\p$ field begins to oscillate, we approximate the dynamics $t \leq t^{\rm A}$ by generalised slow-roll behaviour following eq.~\eqref{eq:attractor}, whereas for $t \geq t^{\rm A}$ we presume it to be a fluid. Both of these approximations break down as $t$ approaches the transition at $t^{\rm A}$. We axiomatically maintain constant energy density over the transition, but the imperfect matching of these two different approximations ensures that we cannot also maintain a constant rate of change of energy density across the transition. We therefore define the oscillation factor $x_\p$ as the ratio of $\d U_\p /  \d N$ and $\d \rho_\p /  \d N$ immediately before and after the onset of oscillations at $t^{\rm A}$. Defining $x_\s$ analogously at $t^{\rm B}$, both oscillation factors may be written as
\begin{equation}
x_\p = \frac{\gpc }{\mpl^2} \frac{3 \, \rho \, \rho_\p}{{U_\p'}^2}\bigg|_{\rm A}
\,, \qquad 
x_\s = \frac{\gsc }{\mpl^2} \frac{c\, \rho \, \rho_\s}{{U_\s'}^2}\bigg|_{\rm B} \,.
\end{equation}
The imperfect nature of the approximations that we make to the true scalar field dynamics therefore appear, not exclusively, in values of $x$ that are deviant from unity. The caveat `not exclusively' is important: even if the density and its first time derivative are continuous over the boundary, there is no guarantee that the higher order derivatives will be, which will prevent perfect matching. For the models considered in this paper we find that these oscillation factors are always of order unity. This suggests that we are not introducing significant error by modelling the scalar field evolution as an initial phase of generalised slow-roll followed by a fluid phase. However, other models may break this condition. Such models may simply require a different choice of oscillation hypersurface, or it is also possible that there may be non-slow roll dynamics that need to be considered.

As an example of how to pick a particular hypersurface, let us consider the prescription of Kawasaki et al.~\cite{Kawasaki:2012gg} where $t^{\rm A}$ is defined by $|\p'/\p|^{\rm A} = 1$ and $t^{\rm B}$ obeys $|\s'/\s|^{\rm B} = 1$. This method seems sensible because it will define oscillations to begin when the field has sufficient velocity to reach the minimum within one efold. One then finds
\begin{subequations}
\begin{equation}
\label{eq:kawasaki}
x_\p = 3 \gpc \frac{U_\p}{\p \, U_\p'}\bigg|_{\rm A} \!, \qquad
x_\s =~ 3 \gsc \frac{U_\s}{\s \, U_\s'}\bigg|_{\rm B} \,, \displaybreak[0]
\end{equation}
\begin{align}
f_{,\s}^{\rm A} &= 1 \,, &
f_{,\p}^{\rm A} =&~ 1 + \frac{\mpl^2}{{\p^{\rm A}}^2} 
\bigg( 1-\frac{{U_\p''} \p}{{U_\p'}} \bigg)_{\rm A} , \displaybreak[0]\\
f_{,\p}^{\rm B} &= 1 \,, & 
f_{,\s}^{\rm B} =&~ 1 + \frac{3 \mpl^2}{c^{\rm B} {\s^{\rm B}}^2} 
\bigg( 1-\frac{{U_\s''}\s}{{U_\s'}} \bigg)_{\rm B} . 
\end{align}
\end{subequations}
These may easily be found for a given potential, for example, if the oscillations begin in the vicinity of a quadratic or quartic potential minimum then:
\begin{itemize}
\item Quadratic: $x = 3/2$ and oscillations begin on a uniform density hypersurface with $f_{,\p}=f_{,\s}=1$.
\item Quartic: $x=1$ and oscillations begin on a non-uniform density hypersurface. At $t^{\rm B}$ one finds $f_{,\p}^{\rm B} = 1$, $f_{,\s}^{\rm B} = 1 - 6 \mpl^2 / (c \s^2)_{\rm B}$, whereas at $t^{\rm A}$ one finds $f_{,\s}^{\rm A} = 1$, $f_{,\p}^{\rm A} = 1 - 2 \mpl^2 / \p^2_{\rm A}$.
\end{itemize}

The final quantities to evaluate are $r^{\rm A}$ and $r^{\rm B}$. These follow simply after we note the following relations
\begin{subequations}
\begin{align}
{\rho_\p'}^{\rm A} &= {U_\p'}^{\rm A} {\p'}^{\rm A} \,, \\
{\rho_\s'}^{\rm A} &= {U_\s'}^{\rm A} {\s'}^{\rm A} \,, \\
{\rho_\p'}^{\rm B} &= -3 \gpc \rho_\p^{\rm B}\,, \\
{\rho_\s'}^{\rm B} &= -3 \mpl^2 {{U_\s'}^{\rm B}}^2 / (c^{\rm B} \rho^{\rm B}) = -3 \gsc \rho_\s^{\rm B} / x_\s \,.
\end{align}
\end{subequations}
Explicit forms for $r^{\rm A}$ and $r^{\rm B}$ the follow as
\begin{align}
\!\!\!\!  \frac{1}{r^{\rm A}} = 1 + \frac{f_{,\p} {U_\p'}^2}{f_{,\s} {U_\s'}^2 } \bigg|_{\rm A} \,, \quad
\frac{1}{r^{\rm B}} = 1 + x_\s \frac{\gpc}{\gsc} \frac{f_{,\p} \rho_\p}{f_{,\s} \rho_\s} \bigg|_{\rm B} .
\end{align}

\subsection{\bf Relating perturbations to horizon exit}
\label{sec:perts}
We now show how the perturbations on the hypersurfaces at $t^{\rm A}$, $t^{\rm B}$, $t^{\rm C}$ and $t^{\rm D}$ can be written in terms of perturbations at horizon exit. \\

\noindent {\bf Perturbations at $\bm{t^{\rm A}}$}:
This relation is equivalent to that derived by Vernizzi and Wands~\cite{Vernizzi:2006ve}. Starting with the equations of motion one can relate the two fields as
\begin{equation}
\label{eq:field_field}
\int_{*}^{\rm A} \! \frac{1}{U_\p'} \, \d \p = 
\int_{*}^{\rm A} \! \frac{1}{U_\s'} \, \d \s \,.
\end{equation}
Varying these integrals yields only four boundary terms and we can eliminate one of $\delta \s^{\rm A}$ or $\delta \p^{\rm A}$ in favour of the other by using eq.~\eqref{eq:arbitrary_gauge} to find
\begin{equation}
\label{eq:Ato*}
\frac{1}{{U_\p'}^{\rm A} r^{\rm A}} \delta \p^{\rm A} = \frac{-1}{{U_\s'}^{\rm A} (1-r^{\rm A})} \delta \s^{\rm A}  =
\frac{\delta \phi^*}{{U_\p'}^*}-\frac{\delta \s^*}{{U_\s'}^*} \,,
\end{equation}
where the choice of hypersurface at $t^{\rm A}$ is arbitrary and is encapsulated within the parameter $r^{\rm A}$. \\

\noindent {\bf Perturbations at $\bm{t^{\rm B}}$}:
This is the most challenging part of our calculation. We seek a relation between the perturbations $\delta \rho_\p^{\rm B}$ to $\delta \rho_\p^{\rm A}$ starting from the relation
\begin{equation}
\label{eq:field_fluid}
\frac{\mpl^2}{c^{\rm B} \gpc} \int_{\rm A}^{\rm B} \! \frac{1}{\rho} \, \d \ln \rho_\p = 
\int_{\rm A}^{\rm B} \! \frac{1}{U_\s'} \, \d \s \,.
\end{equation}
The challenge arises due to the presence of the $\rho$ factor in eq.~\eqref{eq:field_fluid}. There are three cases to consider depending on the relative energy density of the two components throughout phase B:\\

\noindent {\it $\rho_\p$ dominates}: In this case we may approximate $\rho \approx \rho_\p$ and functional variation of eq.~\eqref{eq:field_fluid} then yields only four boundary terms. We can express these in terms of $\delta \rho_\p^{\rm A}$ and $\delta \rho_\p^{\rm B}$ by employing the $f^{\rm A}$ and $f^{\rm B}$ versions of eq.~\eqref{eq:arbitrary_gauge} to eliminate the terms involving $\delta \s^{\rm A}$ and $\delta \s^{\rm B}$. Finally we can write $\delta \rho_\p^{\rm A}$ in terms of horizon exit perturbations using eq.~\eqref{eq:Ato*}. The result is given in eq.~\eqref{eq:Bto*}, where we have defined $\Q^{\rm A}$ and $\Q^{\rm S}$ such that they both equal unity in simple models such as the standard curvaton scenario. \\

\noindent {\it $U_\s$ dominates}: If the $U_\s$ potential dominates the energy density throughout phase B then we may easily adapt the above analytic method to compute $\delta \ln \rho_\p^{\rm B}$ by multiplying eq.~\eqref{eq:field_fluid} by $\rho \approx U_\s$ before integrating. The result then follows by the same procedure as above, arriving at almost identical results. The difference is encapsulated by the modulating parameter $\Q^{\rm S}$. \\

\noindent {\it Codominant case (the inflating curvaton)}: 
The most complex scenario that we may encounter is when neither component dominates throughout phase B. However, even in this case one may make analytic progress by noting that $U_\s$ decays much more slowly than $\rho_\p$. This means that one can split phase B into two sub-phases, the first dominated by $\rho_\p$ and the second dominated by $U_\s$. These two sub-phases are then independently solvable using the above methods, the results of which can then be combined. The second sub-phase drives a second bout of inflation, and has been previously dubbed the `inflating curvaton' model~\cite{Dimopoulos:2011gb,Enomoto:2012uy}. We shall therefore refrain from deriving explicit formulae for this case and instead provide results for the other two scenarios as 
\begin{align}
\label{eq:Bto*}
\frac{\delta \ln \rho_\p^{\rm B}}{x_\s r^{\rm B}} 
&=
\frac{\gpc}{\gsc} \frac{{{U_\s'}^{\rm B}}^2}{U_\s^{\rm B}}
\Q^{\rm S} \Q^{\rm A}
\bigg(\frac{\delta \phi_*}{{U_\p'}^*}-\frac{\delta \s_*}{{U_\s'}^*} \bigg) \,, \\
\Q^{\rm A} &= 1 + \bigg(\frac{3}{c^{\rm B} x_\p} -1 \bigg) r^{\rm A} \,, \\
Q^{\rm S} &= 
\begin{cases}
 1\,, & \text{$\rho_\p$ dominates} \,, \\
 \displaystyle{\frac{U_\s^{\rm A}}{U_\s^{\rm B}}} \,, & \text{$U_\s$ dominates}\,.
\end{cases}
\end{align} 
Note that, even in the case where $U_\s$ dominates, $\Q^{\rm S}$ will still approximately equal unity for any model where phase B is not long enough for $U_\s$ to drop appreciably. \\

\noindent {\bf Perturbations at $\bm{t^{\rm C}}$}:
Perturbations at the beginning and the end of phase C may be related by noting that the background equations of motion provide the identity
\begin{equation}
\int_{\rm B}^{\rm C} \! \frac{1}{\gpc } \, \d \ln \rho_\p = 
\int_{\rm B}^{\rm C} \! \frac{1}{\gsc } \, \d \ln \rho_\s \,.
\end{equation}
Functional variation of this relation yields only four boundary terms.  Using eq.~\eqref{eq:arbitrary_gauge} to eliminate $\delta \rho_\s$ in terms of $\delta \rho_\p$, and then using eq.~\eqref{eq:Bto*} to write $\delta \ln \rho_\p^{\rm B}$ in terms of horizon exit perturbations we find
\begin{align}
\label{eq:Cto*}
\frac{1}{r^{\rm C}} \delta \ln \rho_\p^{\rm C} &= 
\frac{\gpc}{\gsc} \frac{{{U_\s'}^{\rm B}}^2}{U_\s^{\rm B}} 
\Q^{\rm S} \Q^{\rm B} \Q^{\rm A} 
\bigg(\frac{\delta \phi_*}{{U_\p'}^*}-\frac{\delta \s_*}{{U_\s'}^*} \bigg)  \,, \\
\Q^{\rm B} &= 1+ (x_\s-1) r^{\rm B} \,.
\end{align}
Similar to $\Q^{\rm A}$, the modulating parameter $\Q^{\rm B}$ is also unity for simple models such as the standard curvaton. \\

\noindent {\bf Perturbations at $\bm{t^{\rm D}}$}:
For phase D the background equations of motion relate the fluids as
\begin{equation}
\int_{\rm C}^{\rm D} \! \frac{1}{\gpd } \, \d \ln \rho_\p = 
\int_{\rm C}^{\rm D} \! \frac{1}{\gsd } \, \d \ln \rho_\s \,.
\end{equation}
Functional variation yields four boundary terms which we again manipulate using eq.~\eqref{eq:arbitrary_gauge} to eliminate $\delta \rho_\s$ in terms of $\delta \rho_\p$. In this case, the functional form of $r^{\rm C}$ is different to that of $r^{\rm B}$ and so the result takes a correspondingly different form as
\begin{align}
\label{eq:Dto*}
\frac{\delta \ln \rho_\p^{\rm D}}{r^{\rm D}}  &= \frac{\gpd}{\gsd}\Big(
1 + \C_1 r^{\rm C} 
\Big) \frac{{{U_\s'}^{\rm B}}^2}{U_\s^{\rm B}} \Q^{\rm S} \Q^{\rm B} \Q^{\rm A} 
\bigg(\frac{\delta \phi_*}{{U_\p'}^*}-\frac{\delta \s_*}{{U_\s'}^*} \bigg)  \,, \\
\C_1 &=\frac{\gsd \gpc - \gpd \gsc}{\gsc \gpd} \,,
\end{align}
where we have collected some of the $\gamma$ parameters into a constant $\C_1$ where $\C_1=0$ if neither fluid changes equation of state at $t^{\rm C}$. The constant $\C_1$ is therefore zero in simple models such as the standard curvaton model.

\subsection{\bf Putting the pieces together }
\label{sec:dN}

We have now developed all the technology necessary to find $\delta N$ in terms of horizon exit perturbations $\delta \p^*$ and $\delta \s^*$. The terms in the original formula~\eqref{eq:dN_1} for $\delta N$ that are evaluated at $t^{\rm A}$ can be combined and simplified using eq.~\eqref{eq:Ato*} to yield
\begin{align}
& -\frac{U_\p}{\mpl^2 U_\p'}\bigg|_{\rm A} \delta \p^{\rm A}   
- \frac{U_\s}{\mpl^2 U_\s'}\bigg|_{\rm A} \delta \s^{\rm A} 
+\frac{1}{3 \gpc} \delta \ln \rho_\p^{\rm A} = 
\nonumber \\
& \quad 
\frac{1}{\mpl^2} \bigg(\frac{1-x_\p}{x_\p} \rho^{\rm A} r^{\rm A} + U_\s^{\rm A}\bigg)
\bigg(\frac{\delta \phi_*}{{U_\p'}^*}-\frac{\delta \s_*}{{U_\s'}^*} \bigg) \,,
\end{align}
where we have used the relation
\begin{equation}
\frac{1}{3 \gpc} \delta \ln \rho_\p^{\rm A} =
\frac{\rho^{\rm A} r^{\rm A}}{\mpl^2 x_\p} \bigg(\frac{\delta \phi_*}{{U_\p'}^*}-\frac{\delta \s_*}{{U_\s'}^*} \bigg) \,.
\end{equation}
Also substituting for $\delta \ln \rho_\p^{\rm C}$ and $\delta \ln \rho_\p^{\rm D}$ using eqs.~\eqref{eq:Cto*} and~\eqref{eq:Dto*} we obtain our most important result as
\begin{align}
\label{eq:dN_3}
 \delta N &= \bigg[
\frac{U_\s^{\rm A}}{\mpl^2} + \frac{1-x_\p}{\mpl^2 x_\p} \rho^{\rm A} r^{\rm A}
- \Q^{\rm A} \Q^{\rm B} \Q^{\rm C} \Q^{\rm S} 
\frac{r^{\rm D}}{3 \gsd} \frac{{{U_\s'}^{\rm B}}^2}{U_\s^{\rm B}}
\bigg] \nonumber \\
& \qquad \times \bigg(\frac{\delta \phi_*}{{U_\p'}^*}-\frac{\delta \s_*}{{U_\s'}^*} \bigg)  + \delta N_{\rm HCA} \,.
\end{align}
In this formula we have defined a final modulating parameter $\Q^{\rm C}$ 
\begin{align}
\Q^{\rm C} &= 1 + \C_1 r^{\rm C} - \C_2 \frac{r^{\rm C}}{r^{\rm D}} \,, \\
\C_2 &=\frac{\gsd(\gpc-\gpd)}{\gsc \gpd} \,.
\end{align}
One finds $\C_2=0$ if $\rho_\p$ does not change equation of state at $t^{\rm C}$.
This is the case in many simple models such as the standard curvaton model and so one obtains $\Q^{\rm C} = 1$ in this case. 

Eq.~\eqref{eq:dN_3} is the principle results of this paper, providing us with the linear $\delta N$ derivatives $N_{,\p}$ and $N_{,\s}$. This final result demonstrates why we have defined the four modulating parameters $\Q^{\rm A}$, $\Q^{\rm B}$, $\Q^{\rm C}$ and $\Q^{\rm S}$ which are all unity in the the standard curvaton scenario but may be modified in more general circumstances. The $\Q^{\rm A}$ and $\Q^{\rm B}$ account for the effects originating from the transitions at $t^{\rm A}$ and $t^{\rm B}$ respectively. $\Q^{\rm C}$ reflects the changes in the fluid equations of state at $t^{\rm C}$. All three of these $\Q$ parameters also encapsulate any modulation of these hypersurfaces that may be present. Finally, the $\Q^{\rm S}$ parameter accounts for deviations from the `spectator' scenario, where $\rho_\p$ dominates phase B, as we are about to discuss. \\

\noindent {\bf Hypersurface effects}: 
We see that $\delta N$ may be sourced by the transitions occurring at any of the hypersurfaces $t^{\rm A}$, $t^{\rm B}$, $t^{\rm C}$ or $t^{\rm D}$. This effect can be significant if one or many of these transitions occur on a hypersurface that is significantly deviant from one of uniform density whereby the associated $r$-parameters may acquire large magnitudes. Leaving the reheating hypersurfaces general naturally allows us to account for effects such as modulated reheating. The motivation for leaving the oscillation hypersurfaces general is that at present there is no definitive conclusion about which hypersurface one should use for the onset of oscillations. One may perhaps expect that only quite perverse hypersurface choices would have any observational effect, but this is not the case. For example, one may rewrite the standard curvaton result in the form $N_{,\s} \propto \delta \ln \rho_\s^{\rm B}$ and so choosing $t_{\rm B}$ to be a uniform-$\s$ hypersurface would yield $N_{,\s} \simeq 0$. This is clearly incomparable to the standard curvaton result of $N_{,\s} \gg N_{,\p}$, emphasising that the choice of hypersurface can be crucial.\footnote{Note that if we take $t_{\rm B}$ to be a uniform-$\p$ hypersurface then we do not find $N_{,\s} \simeq 0$ because $r^{\rm B} \to 0$ in this same limit.} As a further example, in \S\ref{sec:axion} we shall explicitly show that the condition provided by Kawasaki et al.~\cite{Kawasaki:2012gg} is incomplete.

\subsection{Limits of the general result}
\label{sec:limits}

The general result~\eqref{eq:dN_3} can be applied and simplified to recover a huge range of different scenarios including two-field inflation, the standard curvaton, the modulated curvaton, the spectator scenario, modulated reheating and the inhomogeneous end of inflation scenario. We now show explicitly how four such limits are obtained.\\

\noindent {\bf Inhomogeneous End of Inflation}:
The standard assumption of this scenario is that reheating occurs instantaneously at the end of inflation~\cite{Bernardeau:2002jf,Lyth:2005qk,Salem:2005nd,Alabidi:2006wa,Bernardeau:2007xi,Sasaki:2008uc,Naruko:2008sq}. Thus only phase A is relevant and so the term in eq.~\eqref{eq:dN_3} proportional to $r^{\rm D}$ is not present. In addition, the term $\delta \ln \rho_\p^{\rm A} / 3 \gpc$ in the expressions that lead to eq.~\eqref{eq:dN_3} is also absent since this is generated by integrating the dynamics occurring in phase B. Removing these we find
\begin{align}
\label{eq:iei}
\delta N &= \frac{1}{\mpl^2} \bigg[
U_\s^{\rm A}(1-r^{\rm A}) - U_\p^{\rm A} r^{\rm A}
\bigg] \bigg(\frac{\delta \phi_*}{{U_\p'}^*}-\frac{\delta \s_*}{{U_\s'}^*} \bigg) \nonumber \\
& \qquad + \delta N_{\rm HCA} \,.
\end{align}
If inflation ends on a uniform density hypersurface at $t^{\rm A}$ then $0 \leq r^{\rm A} \leq 1$. One then recovers the HCA result ($\delta N \approx \delta N_{\rm HCA}$) if either field is the dominant energy density at $t^{\rm A}$ or if $\rho^{\rm A} \ll \rho^{\rm *}$. The existence of such limits and how they relate to the onset of adiabaticity is discussed extensively in refs.~\cite{Elliston:2011dr,Elliston:2012wm} and arises due to the dynamical constraints imposed by the assumption of a sum-separable potential. The Inhomogeneous End of Inflation scenario allows $|r^{\rm A}| \gg 1$ which can modify this result, even producing the dominant contribution to $\delta N$ at linear order if the end of inflation hypersurface intercepts the inflationary bundle at a sufficiently large angle~\cite{Elliston:2013afa}.\\

\noindent {\bf Late-Rolling Spectator}:
If a field remains subdominant in energy density throughout inflation then it has little effect on the background dynamics and is often referred to as a `spectator' field. The predictions for spectator scenarios depend significantly on the time at which the spectator field rolls. 

Certain forms of $U_\s$ lead the spectator $\s$ to roll well before the end of inflation. Presuming that the $\s$ field stabilises in some positive mass minimum of its potential then perturbations $\delta \s$ will be attenuated. If such quenching of isocurvature completes before the end of inflation then an adiabatic state is attained by $t^{\rm A}$ and the observational predictions will be determined by the HCA with $\delta N = \delta N_{\rm HCA}$. In this case the post-inflationary dynamics are irrelevant and the formalism of this paper is not required. An explicit model of this type was the axion-quadratic model considered first in Elliston et al.~\cite{Elliston:2011dr} where adiabaticity is reached before the end of inflation if the axion mass is set to be at least five times larger than that of the quadratic inflaton. 

On the other hand, the formalism in this paper is required to consider other choices of $U_\s$ where the $\s$ field rolls at or after the end of inflation.  We now derive the predictions for a late-rolling spectator models where $\s$ rolls significantly after the end of inflation and after the inflaton has reheated to radiation. We therefore take $\gpc = \gpd = 4/3$. This scenario encapsulates the curvaton model, but is significantly more general because we include the inflaton perturbations~\cite{Fonseca:2012cj} and there is no implicit requirement that the $\s$ field perturbations dominate at linear order (i.e. we do not assume $R \gg 1$). We also do not require a specific form of the potential $U_\s$ or a particular equation of state for the effective fluid $\rho_\s$. 

In this case, since $\s$ only rolls well after horizon exit, it is an excellent approximation to consider $\p$ as the adiabatic field at horizon exit and as such it will only receive a constant contribution from the initial boundary at $t^*$. On the other hand, $N_{,\s}$ will be zero initially. To explicitly recover this zero initial condition from full result~\eqref{eq:dN_3} we first note that the above assumptions mean that ${U_\s'}^* \ll {U_\p'}^*$ and so the non-HCA terms in eq.~\eqref{eq:dN_3} will predominantly source $N_{,\s}$ over $N_{,\p}$. Next, since $U_\s^{\rm A} \simeq U_\s^*$, the first term in the square brackets of eq.~\eqref{eq:dN_3} provides the necessary cancellation with the $N_{,\s}$ contribution from $\delta N_{\rm HCA}$. 

The spectator scenario also implies that $r^{\rm A}$ is negligibly small for any hypersurface at $t^{\rm A}$, so long as it obeys $\ep_\s^{\rm A} / \ep_\p^{\rm A} \ll f_{,\p}^{\rm A} / f_{,\s}^{\rm A}$. This allows us to neglect the second term in the square brackets of eq.~\eqref{eq:dN_3} as well as setting $\Q^{\rm A}=1$.\footnote{Extreme hypersurface geometry at $t^{\rm A}$ that does not satisfy $\ep_\s^{\rm A} / \ep_\p^{\rm A} \ll f_{,\p}^{\rm A} / f_{,\s}^{\rm A}$ can be accounted for simply by retaining the second term in the square brackets of eq.~\eqref{eq:dN_3} and the $\Q^{\rm A}$ factor.} Finally, we know that $\Q^{\rm S}=1$ in this case, and so the results become
\begin{subequations}
\begin{align}
\label{eq:spectator_p}
N_{,\p} &\simeq \frac{U_\p}{\mpl^2 U_\p'}\bigg|_* \,, \displaybreak[0]\\
\label{eq:spectator_s}
N_{,\s} &\simeq 
\frac{r^{\rm D}}{3\gsd}   \Q^{\rm C} \Q^{\rm B}
\frac{{{U_\s'}^{\rm B}}^2}{U_\s^{\rm B} {U_\s'}^*} \,.
\end{align}
\end{subequations}
This limit of our calculation is a generalisation of the work of Kawasaki et al.~\cite{Kawasaki:2012gg} who derived a prescription for determining the hypersurface at $t^{\rm B}$ including possible deviations from uniform-density. By ignoring the inflaton perturbations throughout they were implicitly assuming $\Q^{\rm A} = \Q^{\rm C} = 1$. Our result allows for arbitrary hypersurface choices, not only at $t^{\rm B}$, but also at $t^{\rm C}$ and $t^{\rm D}$ (and also at $t^{\rm A}$ by a simple extension). In addition, we include the inflaton perturbations and account for arbitrary equations of state and arbitrary decay order.\\

\noindent {\bf Standard quadratic curvaton}: This is a specific case of the Late-Rolling Spectator limit which occurs where both $U_\p$ and $U_\s$ are of quadratic form and the $\p$ field reheats to radiation before $t^{\rm B}$. One therefore has $\gpc = \gpd = 4/3$ and $\gsc = \gsd = 1$, and since there is no change in the fluid equations of state at $t^{\rm C}$ then $\Q^{\rm C}=1$. The oscillations at $t^{\rm A}$ and $t^{\rm B}$ are assumed to occur on uniform density hypersurfaces and since $r^{\rm A}\ll1$ and $r^{\rm B} \ll1$ we have $\Q^{\rm B} = \Q^{\rm A}=1$. It was to achieve this specific simplification in the standard curvaton case was our motivation for defining the $\Q$ parameters in the form given. We thus recover the standard result
\begin{equation}
\label{eq:standardcurvaton}
N_{,\s} = \frac{2 r^{\rm D}}{3 \s^*}  \,.
\end{equation}

\noindent {\bf Modulated Reheating by a fluid}: 
The conventional modulated reheating setup modulates the hypersurface by allowing it to depend on the value of a slowly rolling field $\s$. This result is not immediately derivable from our four-phase calculation since this requires phase B to be split into two subphases. As with the inflating curvaton, we emphasise that this calculation can easily be done by assembling the ingredients that we have provided in the requisite manner, and we only refrain from providing these details because this particular model has been recently studied elsewhere~\cite{Elliston:2013afa}. Instead, our calculation allows us to consider a new scenario where the reheating hypersurface for the $\p$ field is modulated by its dependence on a fluid density $\rho_\s$. 

The simplest such scenario drives inflation with a single field $\p$ which subsequently oscillates as an effective fluid $\rho_\p$. This fluid then reheats on a hypersurface, the geometry of which is dependent on the value of the $\s$-component, allowing any $\s$-perturbations to alter $\delta N$. In this simplest case, the $\s$ field has a negligible contribution to the energy density throughout and so this is a spectator case. To derive expressions for this model from our framework we will have the $\p$ field reheating at $t^{\rm C}$, reserving $t^{\rm D}$ to be a subsequent uniform density hypersurface such that we may identify $\zeta = \delta N$. Presuming a radiation fluid with $\gpd = 4/3$ after reheating, we require $\gpc \neq 4/3$ to obtain a modification to $\zeta$. Furthermore, the assumptions that the $\s$ field carries no energy density and has negligible linear velocity during inflation means that $r^{\rm A} = r^{\rm B} = r^{\rm D} = 0$ and so $\Q^{\rm S} = \Q^{\rm A} = \Q^{\rm B} = 1$. However, the parameter $r^{\rm C}$ may have a large magnitude if the modulated reheating has significant linear order effects. We therefore find $N_{,\p}$ as in eq.~\eqref{eq:spectator_p} and
\begin{align}
\label{eq:mr_linear}
N_{,\s} &= \frac{(\gpd-\gpc)}{ \gpd} \frac{r^{\rm C}}{3 \gsc} 
\frac{{{U_\s'}^{\rm B}}^2}{U_\s^{\rm B} {U_\s'}^*} \,. 
\end{align}
The factor of $(\gpd-\gpc)$ emphasises the need for the equation of state to change at $t^{\rm C}$ in order for $\zeta$ to be modified. Thus one could get a significant effect at linear order if $|r^{\rm C}| \gg 1$. \\

\section{The Spectral Index and the tensor fraction}
\label{sec:ns}

We now discuss the predictions for $n_\zeta$ and $\tilde r$ which parametrise the linear statistics of inflationary perturbations. Our discussion here is quite general, without requiring a separable potential. 

For this present argument it is helpful to consider a rotated field basis such that $\p$ is the adiabatic field at horizon exit. In this case $N_{,\phi}$ takes a constant value $N_{,\phi}^2 \simeq (2 \ep^* \mpl^2 )^{-1}$ and $N_{,\s}$ is zero at horizon exit but grows subsequently. This means that $R=0$ initially, but can grow to large values at later times. We can then write $n_\zeta$ and $\tilde r$ as
\begin{subequations}
\begin{align}
\tilde r &= \frac{16 \ep^*}{1+R} \,,  \\
\!\!\!\! n_\zeta -1 &= -2 \ep^* + 2 \, \frac{R \, \eta_{\s \s}^* + 2 \sqrt{R} \, \eta_{\p \s}^* + \eta_{\p \p}^* - 2 \ep^*}{1+R} ,
\end{align}
\end{subequations}
where we have included the $\eta_{\p\s}$ term since we are not assuming sum-separability here. 

The effect of the subsequent multifield dynamics on $\tilde r$ is manifestly clear: it decreases. This follows because $R=0$ initially and $R$ is positive definite. The physical explanation of this behaviour is that the multi-field dynamics source additional contributions to $\zeta$, but there is no accompanying growth in the amplitude of gravitational waves. 

The evolution of $n_\zeta$ is almost as simple. There are two limiting values that it reaches in the limits $R \ll1$ and $R \gg 1$, which are
\begin{equation}
n_\zeta -1 = 
\begin{cases}
2 \eta_{\p \p}^* - 6 \ep^*  \,, & \text{for $R \ll 1$ \,,} \\
2\eta_{\s \s}^* - 2\ep^* \,, & \text{for $R \gg 1$ \,.}
\end{cases}
\label{eq:ns_range}
\end{equation}
For a separable potential with $\eta_{\p \s}=0$ then $n_\zeta$ must always lie between these two extremes. For models where $\eta_{\p \s}$ dominates over the other slow-roll parameters then it is possible for $n_\zeta$ to lie outside of the bounds given in eq.~\eqref{eq:ns_range}, but only for intermediate values $R \sim 1$. This behaviour may occur when the field trajectory is undergoing a rapid turn in phase space.

\begin{figure}
\centering
\includegraphics[width=\columnwidth]{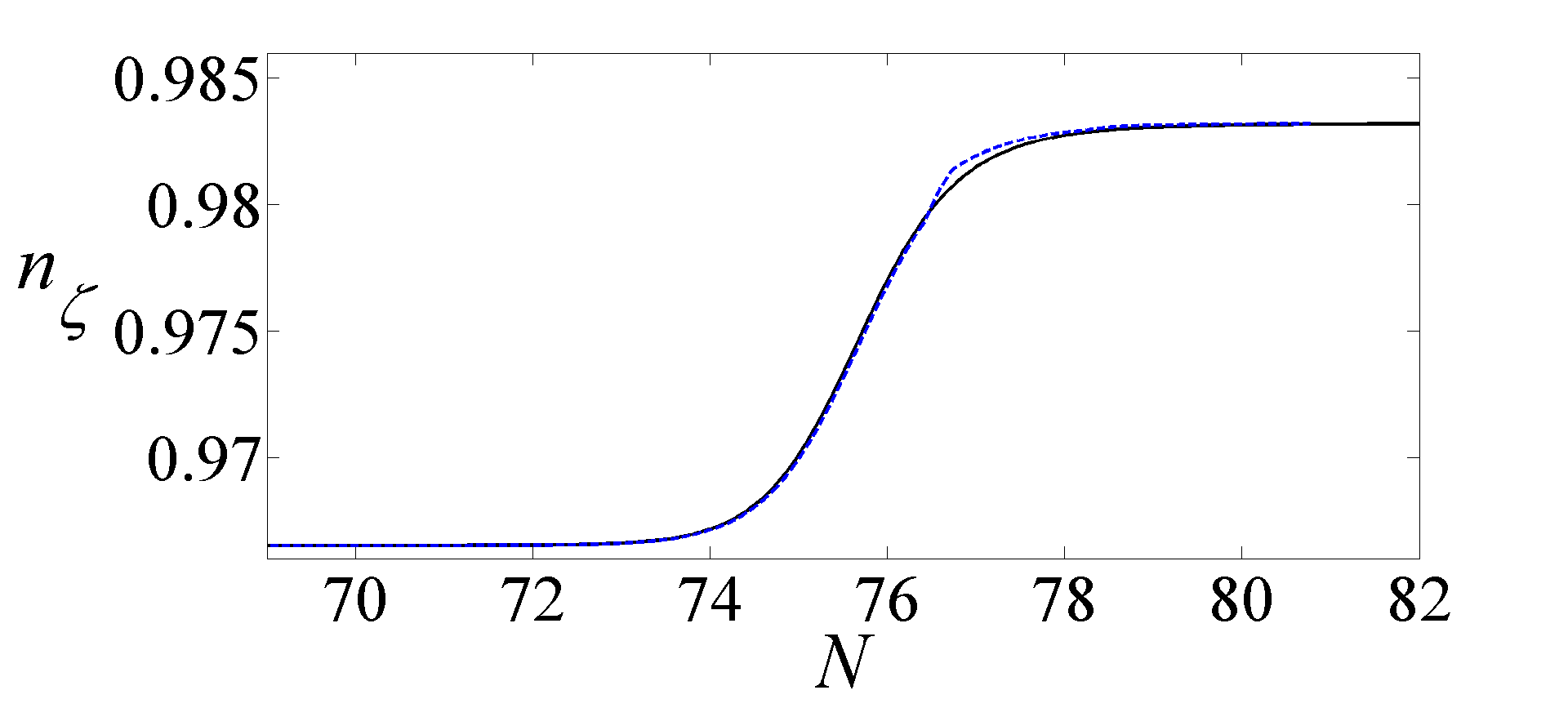}
\caption{Numerical evolution (solid blue line) of $n_\zeta$ for the standard curvaton model ($V = \frac{1}{2} m_\p^2 \p^2 + \frac{1}{2} m_\s^2 \s^2$), plotted against the e-folds $N$. Parameters used are $\s^* = 0.01 \mpl$, $\p^* = 16 \mpl$ and $m_\p = 5 m_\s$ such that $N^{\rm A} \approx 60$. Our analytic method gives the dashed black line. A small value of $\s^*$ is required in order that the curvaton perturbations come to dominate $\zeta$.}
\label{fig:ns}
\end{figure}

Spectator models make a pleasant example of how $n_\zeta$ evolves since the field $\p$ is essentially the adiabatic field at horizon exit. For simple models where the $\s$ perturbations grow monotonically to dominate $\zeta$, there is usually a smooth transition between the two limits given in eq.~\eqref{eq:ns_range} as $R$ increases. When $n_\zeta$ is plotted against $N$, we therefore see a step shape as illustrated for the standard curvaton model in figure~\ref{fig:ns}. In general, there is no guarantee that the $R \gg 1$ regime will be reached, reducing the available range of values that $n_\zeta$ may take. Furthermore, for complex models we should expect non-monotonic evolution of $n_\zeta$ as $R$ fluctuates.

\section{Second order ingredients}
\label{sec:second_order}

The second order derivatives such as $N_{,\s\s}$ follow directly from varying the linear result~\eqref{eq:dN_3}. Writing out the full result is too complex to be intuitively useful, but this is also not desirable because the majority of inflationary scenarios that we would like to consider will not be as general as the most general case studied here. It is therefore more useful for us to provide the ingredients that allow the second order results to be derived, and these may be assembled by the user for the given problem at hand. 

When varying eq.~\eqref{eq:dN_3}, the fundamental ingredients that we need to know are the partial derivatives of the fields or fluids at times $t^{\rm A}$, $t^{\rm B}$, $t^{\rm C}$ and $t^{\rm D}$ with respect to horizon exit field values. These derivatives may respectively and immediately be read from eqs.~\eqref{eq:Ato*},~\eqref{eq:Bto*},~\eqref{eq:Cto*} and~\eqref{eq:Dto*}. One further useful result is the derivative $\partial \s^{\rm B} / \partial \s^*$ which follows from eq.~\eqref{eq:Bto*} as
\begin{equation}
\label{eq:dsds}
\frac{\partial \s^{\rm B}}{\partial \s^*} = 
\frac{{U_\s'}^{\rm B}}{{U_\s'}^*} \big( 1-r^{\rm B} \big) \Q^{\rm S} \Q^{\rm A} \,.
\end{equation}

\noindent {\bf Variations of $\bm{x_\p}$ and $\bm{x_\s}$}: 
Because we do not require any specific definition of the hypersurfaces where oscillations begin, there is no guarantee that the parameters $x_\p$ and $x_\s$ are constant. We therefore provide expressions for their variation below. In some simple situations, however, the $x$-parameters may be very nearly constant, and so such variations are not required. One such example is given by a monomial potential with the oscillation onset condition of Kawasaki et al~\cite{Kawasaki:2012gg} such that $|\s' / \s|=1$ on the hypersurface where the $\s$ field begins to oscillate. But for a general potential or a general hypersurface we find, after eliminating $f_{,\p}$ and $f_{,\s}$ in favour of $r$, that the $x$-parameters vary as
\begin{subequations}
\begin{align}
\frac{\delta x_\p}{x_\p} &= 
\bigg[ 
\bigg( \frac{{U_\p'}^2}{U_\p} - 2 U_\p'' + \frac{{U_\p'}^2}{\rho} \bigg) r
- \frac{{U_\s'}^2}{\rho} \big( 1-r \big)
\bigg]_{\rm A} \times \nonumber \\
& \qquad \bigg( \frac{\delta \p^*}{{U_\p'}^*} - \frac{\delta \s^*}{{U_\s'}^*} \bigg) \,, \\
\frac{\delta x_\s}{x_\s} &= 
\bigg[ 
x_\s \Omega_\p \frac{\gpc }{\gsc } r + 
\bigg( \frac{2 U_\s'' U_\s}{{U_\s'}^2} - 1 - \Omega_\s \bigg) \big( 1-r \big)
\bigg]_{\rm B} \times \nonumber \\
& \qquad \Q^{\rm S} \Q^{\rm A} \frac{{{U_\s'}^{\rm B}}^2}{U_\s^{\rm B}}
\bigg( \frac{\delta \p^*}{{U_\p'}^*} - \frac{\delta \s^*}{{U_\s'}^*} \bigg) \,,
\end{align}
\end{subequations}
from which one may simply read off the partial derivatives such as $\partial x_\p / \partial \s^*$. \\

\noindent {\bf Variations of the $\bm r$-parameters}: 
The variations of $r$ naturally depend upon second derivatives of the hypersurface functions $f$. It proves useful to collect such terms via the dimensionless parameter 
\begin{equation}
\F = \rho_\s \bigg(\frac{f_{,\s\s}}{f_{,\s}} + \frac{f_{,\p\p} f_{,\s}}{f_{,\p}^2} - 2 \frac{f_{,\p\s}}{f_{,\p}}\bigg) \,,
\end{equation}
where $\F = 0$ for a hypersurface of uniform density. We then find the variations of the $r$-parameters as
\begin{subequations}
\begin{align}
\delta r^{\rm A} &= -
\bigg[ \bigg(\F \frac{{U_\s'}^2}{U_\s} + 2 U_\s'' \bigg) \big(1-r \big) + 2 U_\p'' r
\bigg]_{\rm A} \nonumber \\
& \qquad \times r^{\rm A} \big(1-r^{\rm A}\big) 
\bigg( \frac{\delta \p^*}{{U_\p'}^*} - \frac{\delta \s^*}{{U_\s'}^*} \bigg) \,, 
\displaybreak[0]\\
\delta r^{\rm B} &= - 
\bigg[ \bigg( 
\F - \Omega_\s + 2\frac{U_\s'' U_\s}{{U_\s'}^2} \bigg) 
\big( 1-r \big) 
+ x_\s \frac{\gpc}{\gsc} \big(1 + \Omega_\p \big) r
 \bigg]_{\rm B} \nonumber \\
& \qquad \times r^{\rm B} \big(1-r^{\rm B}\big) \Q^{\rm A} \Q^{\rm S}
\frac{ {{U_\s'}^{\rm B}}^2}{U_\s^{\rm B}}
\bigg( \frac{\delta \p^*}{{U_\p'}^*} - \frac{\delta \s^*}{{U_\s'}^*} \bigg)\,, 
\displaybreak[0]\\
\label{eq:deltarc}
\delta r^{\rm C} &= - 
\bigg[ \big( 1 + \F \big) \big( 1-r \big) 
+ \frac{\gpc}{\gsc} r
 \bigg]_{\rm C} \nonumber \\
& \qquad \times r^{\rm C} \big(1-r^{\rm C}\big) \Q^{\rm A} \Q^{\rm B} \Q^{\rm S}
\frac{{{U_\s'}^{\rm B}}^2}{U_\s^{\rm B}} 
\bigg( \frac{\delta \p^*}{{U_\p'}^*} - \frac{\delta \s^*}{{U_\s'}^*} \bigg) \,, 
\displaybreak[0]\\
\delta r^{\rm D} &= - 
\bigg[ \big( 1 + \F \big) \big( 1-r \big) 
+ \frac{\gpd}{\gsd} r
 \bigg]_{\rm D} \Big( 1 + \C_1 r^{\rm C} \Big) \nonumber \\
& \qquad \times r^{\rm D} \big(1-r^{\rm D}\big) \Q^{\rm A} \Q^{\rm B} \Q^{\rm S}
\frac{{{U_\s'}^{\rm B}}^2}{U_\s^{\rm B}} 
\bigg( \frac{\delta \p^*}{{U_\p'}^*} - \frac{\delta \s^*}{{U_\s'}^*} \bigg) \,.
\end{align}
\end{subequations}
These general results account for second order effects arising from the modulated transition at any of the hypersurfaces at $t^{\rm A}$, $t^{\rm B}$, $t^{\rm C}$ or $t^{\rm D}$. \\

\noindent {\bf Uniform density case:} 
In this scenario the above results simplify considerably since $f_{,\p} = f_{,\s} = 1$ and $\F=0$. One finds
\begin{subequations}
\begin{align}
\frac{\delta x_\p}{x_\p} &= 
\bigg[ 
\frac{{U_\p'}^2}{U_\p} - 2 U_\p'' 
\bigg]_{\rm A} r^{\rm A} \bigg( \frac{\delta \p^*}{{U_\p'}^*} - \frac{\delta \s^*}{{U_\s'}^*} \bigg) \,, \\
\frac{\delta x_\s}{x_\s} &= 
\bigg[ 
\frac{{U_\s'}^2}{U_\s} - 2 U_\s'' 
\bigg]_{\rm B} \big(r^{\rm B} -1 \big)\Q^{\rm S} \Q^{\rm A} 
\bigg( \frac{\delta \p^*}{{U_\p'}^*} - \frac{\delta \s^*}{{U_\s'}^*} \bigg) \,,
\end{align}
\end{subequations}
and we clearly see that these are zero for quadratic potentials $U_\p$ and $U_\s$. The $r$ parameters also simplify in the uniform density case as
\begin{subequations}
\begin{align}
\delta r^{\rm A} &= \bigg[
2U_\s'' \big( 1-r \big) + 2U_\p'' r
 \bigg]_{\rm A} \nonumber \\
 & \qquad \times r^{\rm A} \big(r^{\rm A}-1\big) 
\bigg( \frac{\delta \p^*}{{U_\p'}^*} - \frac{\delta \s^*}{{U_\s'}^*} \bigg), \\
\delta r^{\rm B} &= \Q^{\rm A} \Q^{\rm S}
\bigg[ 2 {U_\s''}^2 + \bigg( x_\s \frac{\gpc {U_\s'}^2}{\gsc U_\s} - 2 {U_\s''} \bigg) r 
\bigg]_{\rm B} \nonumber \\
& \qquad \times r^{\rm B} \big(r^{\rm B}-1\big) 
\bigg( \frac{\delta \p^*}{{U_\p'}^*} - \frac{\delta \s^*}{{U_\s'}^*} \bigg)\,, \\
\delta r^{\rm C} &= \Q^{\rm A} \Q^{\rm B} \Q^{\rm S}
\bigg[ 1 + \frac{\gpc-\gsc}{\gsc} r^{\rm C}
 \bigg] \nonumber \\
&  \qquad \times  r^{\rm C} \big(r^{\rm C}-1\big)
\frac{{{U_\s'}^{\rm B}}^2}{U_\s^{\rm B}} 
\bigg( \frac{\delta \p^*}{{U_\p'}^*} - \frac{\delta \s^*}{{U_\s'}^*} \bigg) \,, \\
\delta r^{\rm D} &= \Q^{\rm A} \Q^{\rm B} \Q^{\rm S}
\big( 1 + \C_1 r^{\rm C} \big) \bigg[ 1 + \frac{\gpd-\gsd}{\gsd} r^{\rm D} \bigg] 
 \nonumber \\
& \qquad \times r^{\rm D} \big(r^{\rm D}-1\big) 
\frac{{{U_\s'}^{\rm B}}^2}{U_\s^{\rm B}} 
\bigg( \frac{\delta \p^*}{{U_\p'}^*} - \frac{\delta \s^*}{{U_\s'}^*} \bigg) \,.
\end{align}
\end{subequations}

\noindent {\bf Interpretation: }
At linear order, the necessary detail about the geometry of a given hypersurface is encapsulated in the parameter $r$. At second order we need also to consider $\F$. In the uniform density case, $r$ is bounded as $0 \leq r \leq 1$ where $r=0$ corresponds to a dominant $\p$ component and $r=1$ occurs when the $\s$ component dominates. In either of these limits we observe that $\delta r = 0$. For uniform density hypersurfaces we therefore only expect modulation effects from $\delta r$ in the intermediate regime. At $t^{\rm B}$, $t^{\rm C}$ and $t^{\rm D}$ this corresponds to the scenario where both fields have a non-negligible contribution to the total energy density, whereas at $t^{\rm A}$ this corresponds to the case where both fields are moving with comparable velocities (for most simple potentials this would correspond to the scenario where both fields begin oscillations at about the same time). These conditions make intuitive sense, because if one field totally dominates then the phase space dynamics are essentially linear motion; it is only in the presence of turning that isocurvature modes are able to modify $\zeta$.

For the non-uniform density case, however, this condition is relaxed. One still requires $r$ to deviate from zero or unity for any significant effect to occur, but in this case $r$ may take any real value depending on the geometry of the hypersurface in question. This makes it considerably easier for large values of $\delta r$ to be achieved.

\section{Bispectra in specific cases}
\label{sec:models}
We now consider the bispectrum parameter $\fnl$ in a few interesting limiting cases. First, to recover and extend previous results, we shall consider Late-Rolling Spectator models for a general potential $U_\s$ but without modulation at $t^{\rm B}$. A specific limit this result recovers the standard curvaton scenario, without neglecting the inflaton perturbations. As an illustration of how different forms for $U_\s$ can alter this result, we shall discuss a simple hilltop potential. Next, we shall test the analytic criterion of Kawasaki et al.~\cite{Kawasaki:2012gg} for defining the hypersurface on which oscillations begin. We will do this with the axion quadratic model~\cite{Elliston:2011dr,Leung:2012ve} as an example of a spectator model where oscillations begin on a modulated hypersurface. In doing so we will show that our analytic formalism is able to explain the numerical results of ref.~\cite{Leung:2012ve}. Finally, we shall use the technology introduced in this paper to consider a novel scenario of modulated reheating where the modulation is determined by a fluid rather than a field.

In all cases, a bispectrum analysis requires that we compute the linear perturbation ratio $R$, as well as the ratios of $\delta N$ coefficients $N_{,\s\s} / N_{,\s}^2$, $N_{,\p\p} / N_{,\p}^2$ and $N_{,\p\s} / N_{,\p} N_{,\s}$. \\

\subsection{Late-Rolling Spectator models}
\label{sec:spectator}
This scenario was introduced in \S\ref{sec:limits} where it was shown how to calculate its linear order statistics. Our aim in this section is to show how the parameter $\fnl$ may be a sensitive probe of the particular form of the potential $U_\s$, thereby eliminating some regions of potential space. 

We make the same assumptions as before. The inflaton reheats before the $\s$ field rolls, giving $\gpc = \gpd = 4/3$. We choose $\rho_\s$ to reheat at $t^{\rm D}$ and so $\gsc=\gsd$ and $\Q^{\rm C}=1$. In addition, to simplify the present discussion, we shall restrict to scenarios where $\Q^{\rm B} = 1$. For the spectator cases this does not place significant constraints on the hypersurface at $t^{\rm B}$ because one only requires 
$f_{,\s}^{\rm B} / f_{,\p}^{\rm B} \ll x_\s \rho_\p^{\rm B} / \rho_\s^{\rm B}$ where the right-hand-side of this inequality is very large in the spectator scenario. We can then find $N_{,\p\p}$ and $N_{,\s\s}$ by taking the derivatives of $N_{,\p}$ and $N_{,\s}$ as given in eqs.~\eqref{eq:spectator_p} and eq.~\eqref{eq:spectator_s} to find
\begin{subequations}
\begin{align}
\label{eq:Rspec}
\sqrt{R} &= \frac{\mpl^2 {r^{\rm D}}}{3 \gsd} \frac{{{U_\s'}^{\rm B}}^2}{{U_\s^{\rm B}} {{U_\s'}^*}}
\frac{{{U_\p'}^*}}{{U_\p^*}} \,, \displaybreak[0]\\
\label{eq:Npp_spec}
\frac{N_{,\p\p}}{N_{,\p}^2} &= \mpl^2 \bigg( \frac{{U_\p'}^2}{U_\p^2} - \frac{U_\p''}{U_\p} \bigg)_* \,, \displaybreak[0]\\
\frac{N_{,\s\s}}{N_{,\s}^2} &= \big(3 \gsd-4\big) r^{\rm D} +
4-6\gsd
\nonumber \\ \label{eq:Nss_spec}
& \qquad \qquad + \frac{3 \gsd}{r^{\rm D}} \frac{U_\s^{\rm B}}{{{U_\s'}^{\rm B}}^2} \big(2 {U_\s''}^{\rm B} -{U_\s''}^*\big) \,,
\end{align}
\end{subequations}
and we have not included the $N_{,\p\s}$ term since we find this to provide a subdominant contribution to $\fnl$. We note that the final term in eq.~\eqref{eq:Nss_spec} has the form $(2 \eta_{\s\s}^{\rm B} - \eta_{\s\s}^*)$ which is of the same form as the slow-roll contributions to $\fnl$ as identified in eq.~(3.8) of ref.~\cite{Elliston:2012wm} under the identification of $\s$ as the isocurvature field. 

For times $t \leq t^{\rm B}$ the inflaton perturbations dominate and $\fnl \sim \O(\ep^*)$. Later on, as $r^{\rm D}$ grows, then it is {\it possible} that $R$ will grow to exceed unity. Presuming this to be the case, then the $N_{,\s\s}$ term will now contribute to $\fnl$. For small $r^{\rm D}$ (but not so small that $R \ll 1$) the final term in eq.~\eqref{eq:Nss_spec} is dominant. Since $t^{\rm D}$ is a uniform density hypersurface, $r^{\rm D}$ saturates at a value of unity in the limit where the $\s$ field dominates the energy density. Presuming $R \gg 1$ at the same time, which is not always the case, one finds $\fnl$ takes a limiting value
\begin{equation}
\label{eq:fnl_spect_lim}
\fnl \to \frac{5 \gsd}{2} \bigg[ \frac{U_\s^{\rm B}}{{{U_\s'}^{\rm B}}^2} \big(2 {U_\s''}^{\rm B} -{U_\s''}^* \big) -1 \bigg] \,.
\end{equation}
Therefore, whether we are in the limit $r^{\rm D} \ll 1$ or $r^{\rm D} \to 1$, we see that $\fnl$ is enhanced for any model where $U_\s U_\s'' \gg {U_\s'}^2 $. Such models, which includes the hilltop case we are shortly to consider, are therefore ruled out by Planck bounds of $\fnl = 2.7 \pm 5.8$ at $68$\%~C.L.~\cite{Ade:2013ydc}. 

\subsection{Double Quadratic}
\label{sec:dquad}
\begin{figure}[h]
\centering
\includegraphics[width=\columnwidth]{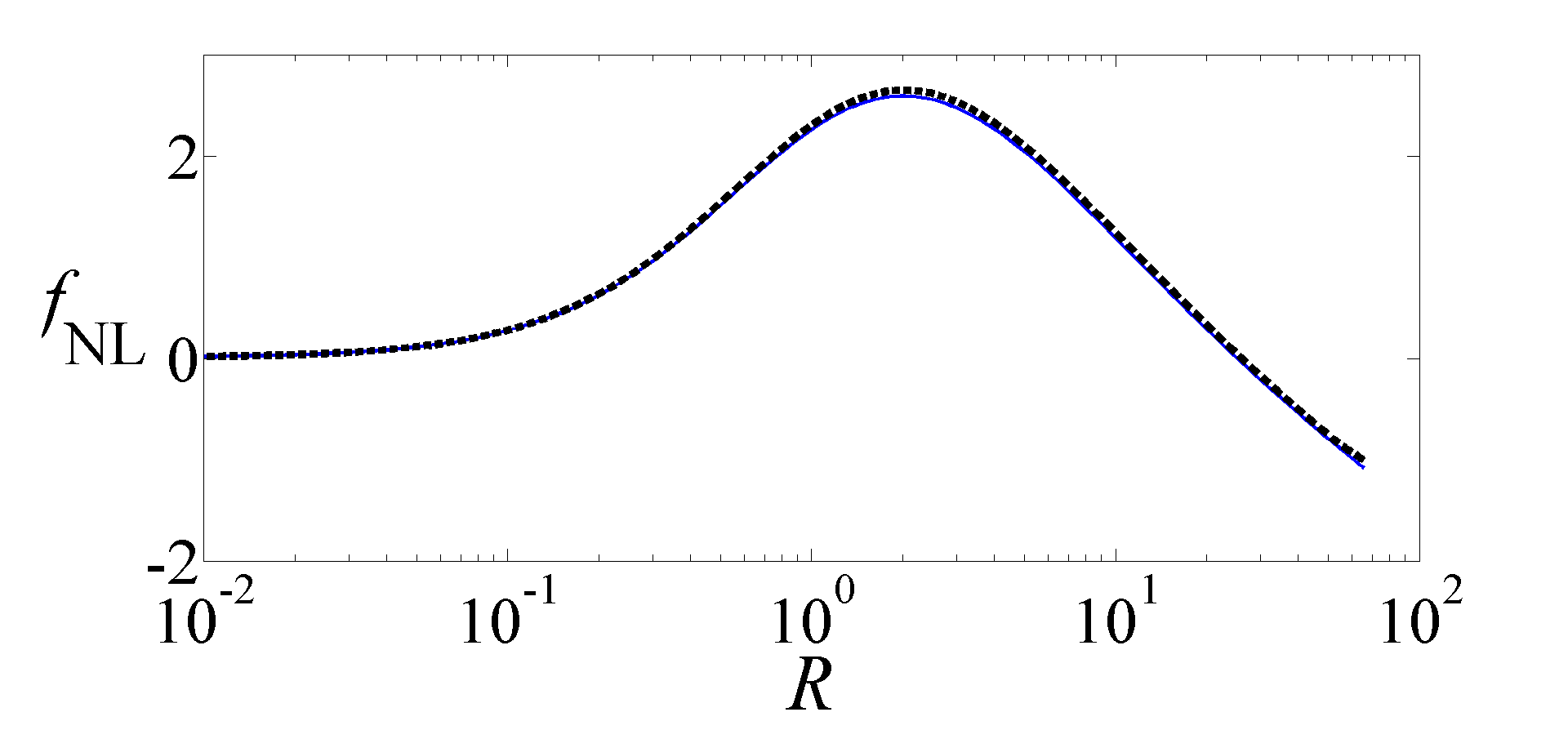}
\caption{Numerical evolution (solid blue line) of $\fnl$ for the standard double quadratic curvaton model, plotted against $R$. Parameters used are $\s^* = 0.01 \mpl$, $\p^* = 16 \mpl$ and $m_\p = 5 m_\s$ such that $N^{\rm A} \approx 60$. Our analytic method gives the dashed black line. Consistent with the analytic predictions, $\fnl$ begins with a small value, before increasing to a positive peak near to $R \approx 3$ with $\fnl \approx 2.3$ at the peak itself. For larger values of $R$ the value of $\fnl$ reduces towards the limit $\fnl \to -5/4$.}
\label{fig:curvaton}
\end{figure}

We include this model only to demonstrate that we recover this standard result. Setting $U_\p = \frac{1}{2} m_\p^2 \p^2$ and $U_\s = \frac{1}{2} m_\s^2 \s^2$ and also $\gsc = \gsd = 1$ in eqs.~\eqref{eq:Rspec}-\eqref{eq:Nss_spec} yields
\begin{subequations}
\begin{align}
\label{eq:R_curv}
\sqrt{R} &= \frac{2 r^{\rm D}}{3 \sqrt{N^{\rm A}}} \frac{\mpl}{\s^*}\,, \\
\frac{N_{,\p\p}}{N_{,\p}^2} &= \frac{1}{2N^{\rm A}} \,, \\
\frac{N_{,\s\s}}{N_{,\s}^2} &= -r^{\rm D} -2 +\frac{3}{2 r^{\rm D}} \,,
\end{align}
\end{subequations}
where $N^{\rm A} \approx {\p^*}^2 / 4 \mpl^2$ is the number of efolds of inflation. In the limit where the inflation perturbations are neglected, these results recover the curvaton result of Lyth and Rodr\'{i}guez~\cite{Lyth:2005fi}. Again, $0 \leq  r^{\rm D} \leq 1$ since $t^{\rm D}$ is a uniform density hypersurface.  Before the curvaton has any effect on the linear perturbations then $R \ll 1$ and $\fnl$ assumes a negligible single-field value of $\fnl \simeq 5 / (12 N^{\rm A}) \sim 10^{-2}$. The $\s$ perturbations can come to dominate the linear order statistics $(R \gg 1)$ at later times, but from the form of eq.~\eqref{eq:R_curv} we can immediately see that this is only possible for models with $\s^* \ll \mpl$. Presuming that $R$ does indeed grow to take a value significantly exceeding unity, one finds that $\fnl$ increases to a positive peak at $R \approx 3$ where $\fnl \approx 0.035 \, \mpl / \s^*$ before settling down to an asymptotic value in the limit of large $R$ where $\fnl = -5/4$. Planck constraints on $\fnl$ therefore eliminate initial field values in close proximity to the minimum of this potential. We plot the dependence of $\fnl$ against $R$ in figure~\ref{fig:curvaton}, where we highlight the fact that $\fnl$ is small for small $R$ because the inflaton perturbations dominate in this regime. 

\subsection{Hilltop curvaton}
\label{sec:hilltopcurvaton}
Moving beyond quadratic forms for $U_\s$ or $U_\p$ will clearly modify the observational predictions. An interesting scenario are `hilltop' potentials where $\s^*$ lies in a plateau region of the potential $U_\s$, such as models where $\s$ is an axion and $U_\s$ assumes a sinusoidal form. For simplicity we keep $U_\p$ as a standard quadratic potential. In order to keep the discussion somewhat generic we define the hilltop form of $U_\s$ by using the lowest order Taylor expansion which yields a quadratic term in the simplest of cases. An effective hilltop potential can therefore be written as $U_\s = {\rm const.} - \frac{1}{2} m_\s^2 (\s-\s_0)^2$ where we take the constant to equal $\frac{1}{2}m_\s^2 \s_0^2$. Clearly this potential has strange behaviour as $\s \to 0$, but this is not an issue because the field will begin to oscillate as a fluid much closer to the hilltop. 

To determine the predictions, we need to know $\s^{\rm B}$. We calculate this from the background equations of motion by presuming the oscillation condition of Kawasaki et al.~\cite{Kawasaki:2012gg}. We also presume a large mass ratio $m_\p^2 \gg m_\s^2$ in order to realise the Late-Rolling condition. The calculation of $\s^{\rm B}$ is placed in the appendix. The result eq.~\eqref{eq:sig_hilltop} gives $(\s_0-\s^{\rm B}) \simeq e^{1/4} (\s_0 - \s^*)$ for $\gamma_\p = 4/3$, informing us that the field does not roll far before it begins to oscillate. 

$N_{,\p\p}/N_{,\p}^2$ follows identically to the standard curvaton model, but $R$ and $N_{,\s\s}/N_{,\s}^2$ are modified. In the hilltop limit where $(\s_0-\s^*) \ll \s_0$ we find
\begin{subequations}
\begin{align}
\sqrt{R} &\simeq \frac{2 r^{\rm D} e^{1/2}}{3 \gsd \sqrt{N^{\rm A}}}  \frac{\mpl}{\s_0} \frac{\s_0-\s^*}{\s_0}\,, \\
\frac{N_{,\s\s}}{N_{,\s}^2} &\simeq (3 \gsd-4) r^{\rm D} +
4-6\gsd  - \frac{3 \gsd}{2 r^{\rm D}}  
\frac{\s_0^2}{(\s_0-\s^{\rm B})^2}  \,,
\end{align}
\end{subequations}
and away from the hilltop limit one finds results qualitatively similar to those of the standard double quadratic. In the hilltop limit we see that $R$ will always be small unless $U_\s$ is a small-field potential with $\s_0 \ll \mpl$, and the initial condition on $\s^*$ must not be too close to the eternal inflation point at the apex of the hilltop. Provided that these conditions are met, then we see that this choice of potential $U_\s$ has a greater capacity to generate large $\fnl$, with larger values arising the closer $\s^*$ is to the hilltop maximum.

\subsection{Quadratic Axion} 
\label{sec:axion}
We now consider the quadratic axion model of Elliston et al.~\cite{Elliston:2011dr}. We will now demonstrate that our analytic formalism is capable of describing the numerical behaviour found in Leung et al.~\cite{Leung:2012ve} for this model. A second aim here is to investigate the validity of the criterion for the onset of oscillations that is provided by Kawasaki et al.~\cite{Kawasaki:2012gg}. We have singled out this particular prescription because it appears sensible and has also been shown to give accurate results previously. The criterion suggests that oscillations begin on a hypersurface defined by $|\s' / \s| = \mu$, where the authors of ref.~\cite{Kawasaki:2012gg} claim that the constant $\mu$ is of order unity, but that observables are independent of its precise value. 

Their work considered only the spectator case. Taking this limit of our results, and presuming that any modulation at times $t^{\rm A}$ and $t^{\rm B}$ is sufficiently small, then any explicit dependence on the $x$-parameters drops out. In this limit we therefore agree that the value of $\mu$ should have no bearing on the observational predictions. However, as one moves away from the spectator limit then we anticipate different behaviour. We therefore consider a model that is not a spectator model, and investigate the consequences of varying $\mu$. 

\begin{figure}[h]
\centering
\includegraphics[width=\columnwidth]{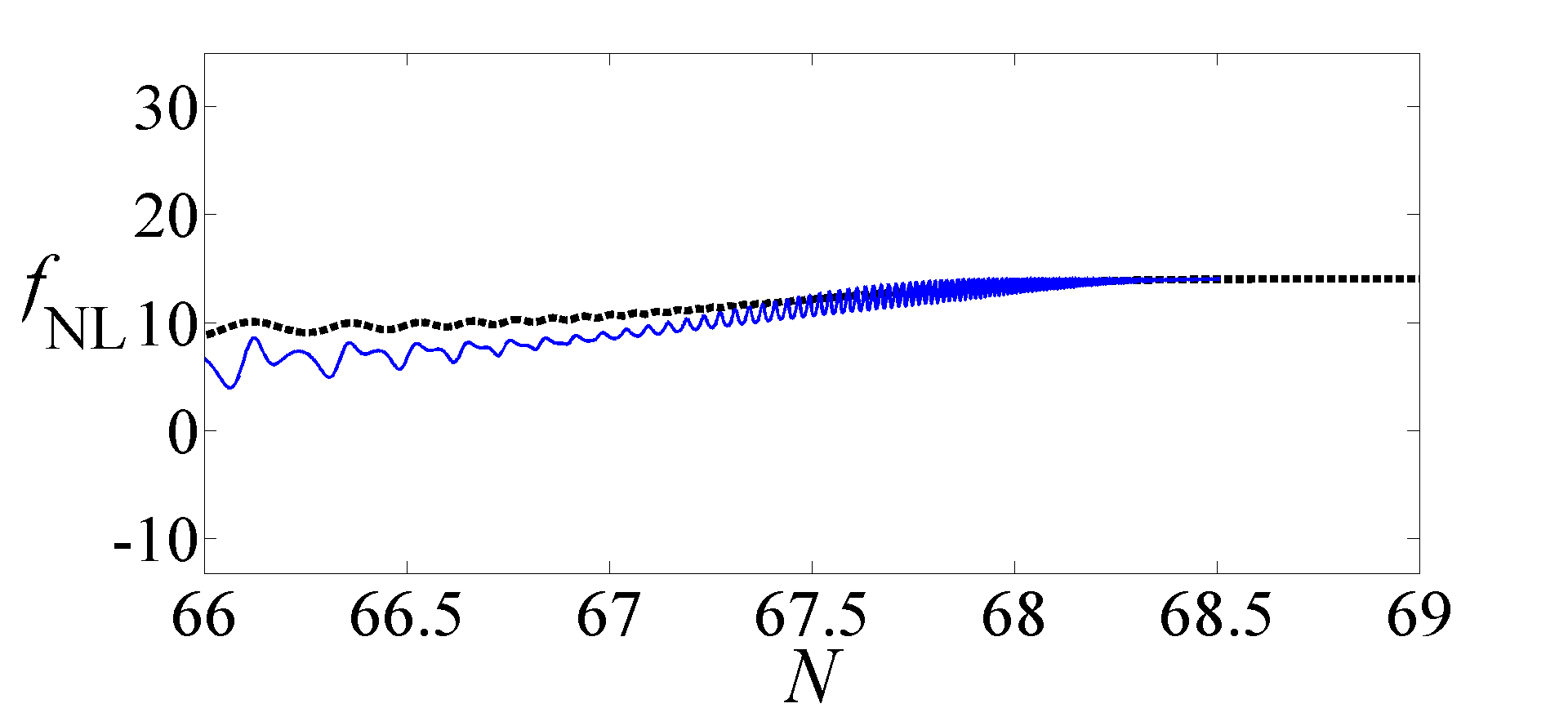}
\includegraphics[width=\columnwidth]{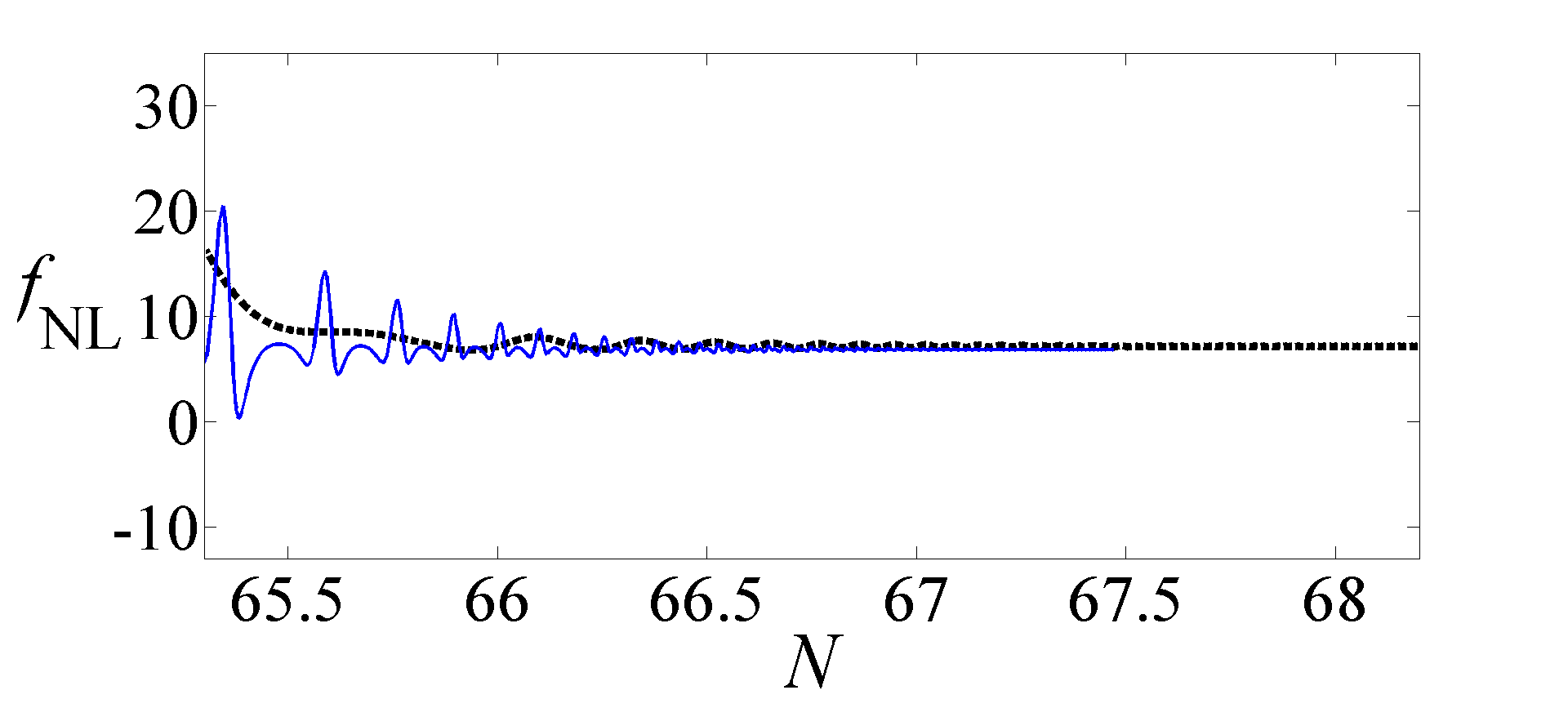} 
\includegraphics[width=\columnwidth]{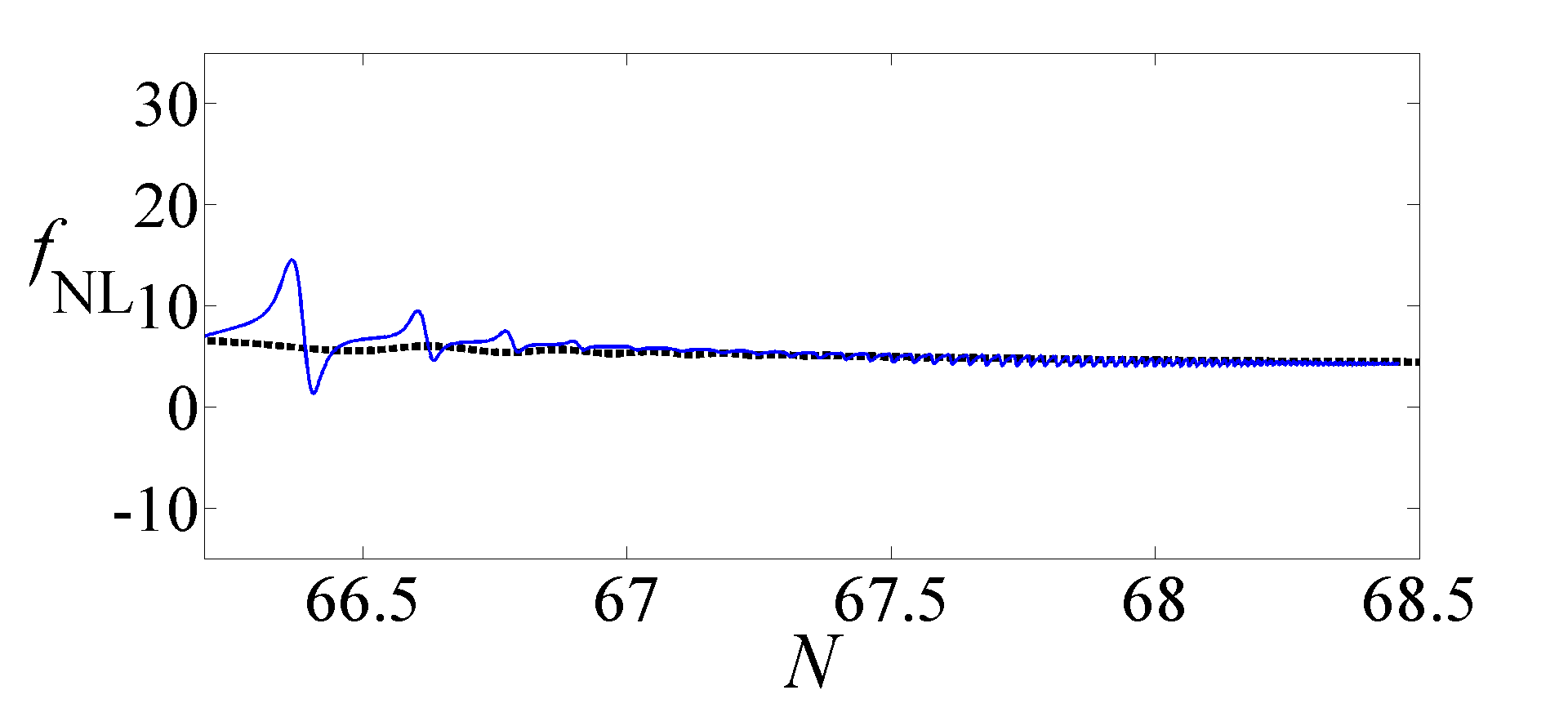} 
\caption{Plots of $\fnl$ for the axion quadratic model with parameters as given under eq.~\eqref{ax}. The solid blue line is the numerical result and the dashed black line is the analytic result. For the top pane we use $\mu=0.6$, and the other two panes use $\mu=0.8$. The values of the $\Gamma$ parameters, in units of $m_\p$, are: Top pane: $\Gamma_\phi=0.01$ and $\Gamma_\sigma=0.1$, Middle pane: $\Gamma_\phi=0.01$ and $\Gamma_\sigma=0.01$, Bottom pane: $\Gamma_\phi=0.1$ and $\Gamma_\sigma=0.01$.
}
\label{fig:axion}
\end{figure}

For the axion quadratic model the $\p$ field follows the quadratic potential $U_\p = \frac{1}{2} m_\p^2 \p^2$ and $U_\s$ takes the form
\begin{equation}
U_\s = \Lambda^4 \bigg( 1-\cos \Big( \frac{2 \pi \s}{f} \Big)\bigg) \,,
\label{ax}
\end{equation}
where $0 \leq \s \leq f/2$. We take parameters of $f=1$, $\Lambda^4 = m_\p^2 f^2 /(4 \pi^2)$, $\p^* = 16 \mpl$ and $\s^* = (\frac{f}{2}-0.001) \mpl$. We then vary the values of the $\Gamma$ parameters. These are all illustrated in figure \ref{fig:axion} where we see that the analytic method is providing a good fit to the numerical behaviour. This demonstrates that the numerical results of Leung et al.~\cite{Leung:2012ve} are simply explained by considering the relative redshifting of the two fluids involved, consistent with the conclusions of Meyers and Tarrant~\cite{Meyers:2013gua}.

\begin{figure}[h]
\centering
\includegraphics[width=\columnwidth]{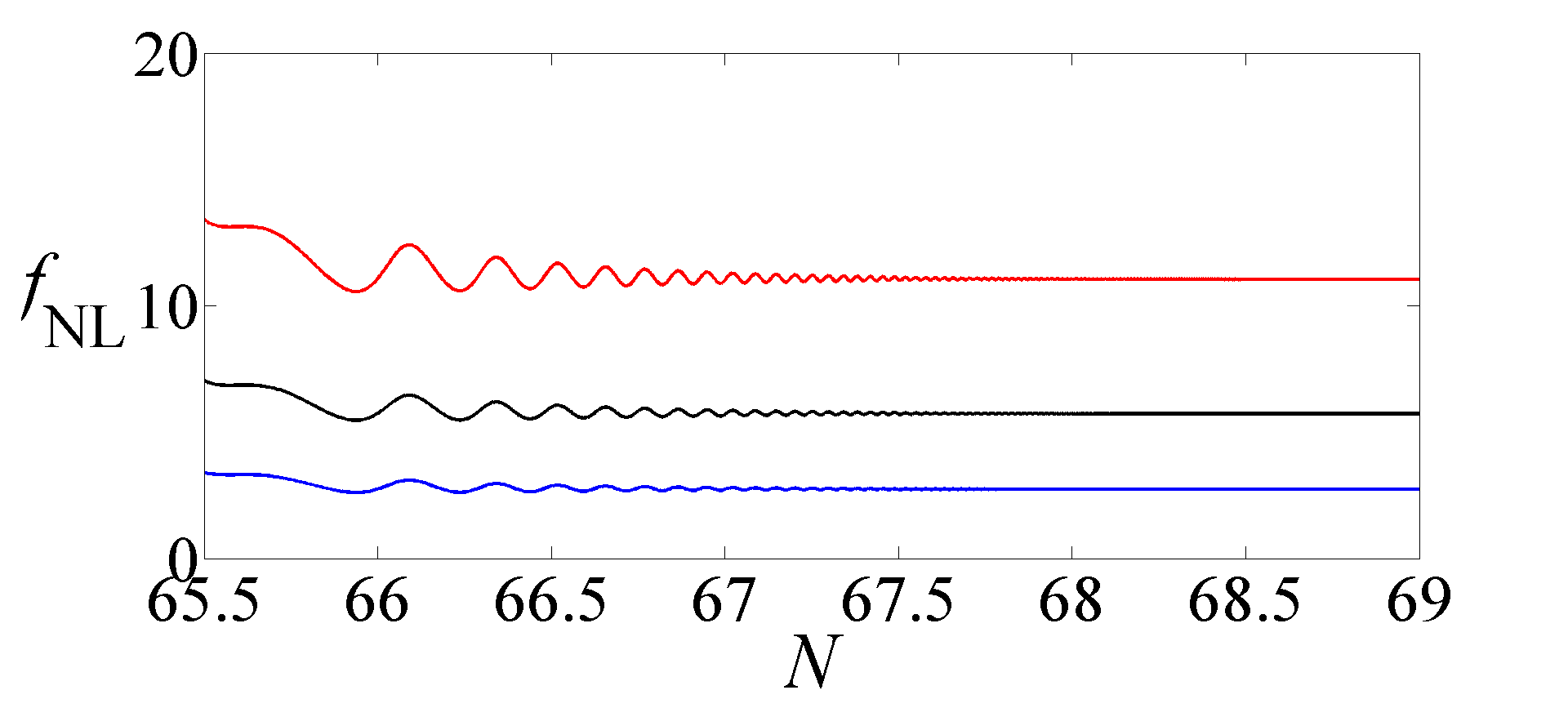}
\caption{Plot of the analytic evolution of $\fnl$ for the axion quadratic model with parameters as given under eq.~\eqref{ax}. The decay rates are $\Gamma_\phi=0.01$ and $\Gamma_\sigma=0.01$. The red, black and blue lines correspond to respectively $\mu=0.5$, $\mu=1$ and $\mu=2$, showing that the analytic estimates depend quite sensitively on the exact value of $\mu$. This demonstrates a need for a better analytic condition for the onset of oscillations.}
\label{fig:axmu}
\end{figure}

In figure ~\ref{fig:axmu} we show a plot of $\fnl$, calculated analytically for three different values of $\mu$. The fact that the analytic results are so sensitive to the parameter $\mu$ motivates a need for better analytic methods to describe when a scalar field begins to oscillate. Our analytic formalism can be used as a testing ground for such analytic methods because it applies equally to all methods for defining the hypersurface of oscillation onset.

\subsection{Modulated reheating of a fluid, by a fluid} 

Let us now consider a new scenario where the decay of the $\p$ field is modulated by a fluid $\rho_\s$. This setup was described in the text leading to eq.~\eqref{eq:mr_linear}, which combined with eq.~\eqref{eq:spectator_p} gives the linear $\delta N$ coefficients. In order to differentiate these we note that since $t^{\rm C}$ is not a uniform density hypersurface then $\delta r^{\rm C}$ follows from eq.~\eqref{eq:deltarc}. To differentiate quantities depending on $\s^{\rm B}$ we also employ eq.~\eqref{eq:dsds} where in the present scenario $\Q^{\rm S} = \Q^{\rm A} = 1$ and $r^{\rm B}$ is negligible. For a general potential $U_\s$ we then find
\begin{subequations}
\begin{align}
\sqrt{R} &= \frac{\mpl^2 {r^{\rm C}}}{3 \gsc}
\frac{(\gpd-\gpc)}{ \gpd}
\frac{{{U_\s'}^{\rm B}}^2}{{U_\s^{\rm B}} {{U_\s'}^*}}
\frac{{{U_\p'}^*}}{{U_\p^*}} \,, \\
\label{eq:NppMR}
\frac{N_{,\p\p}}{N_{,\p}^2} &= \mpl^2 \bigg( \frac{{U_\p'}^2}{U_\p^2} - \frac{U_\p''}{U_\p} \bigg)_*
=2 \ep_\p^* - \eta_{\p\p}^* \,, \\
\frac{N_{,\s\s}}{N_{,\s}^2} &= 
\frac{3\gsc \gpd}{\gpd - \gpc} \bigg[
\F^{\rm C} \frac{(1-r^{\rm C})^2}{r^{\rm C}} + r^{\rm C} - 2 
+ \frac{\gpc}{\gsc} (1-r^{\rm C}) \nonumber \\ 
& \qquad 
+\frac{U_\s^{\rm B}}{{{U_\s'}^{\rm B}}^2 r^{\rm C}} 
\bigg( 2 {U_\s''}^{\rm B} - {U_\s''}^* \bigg)
\bigg] \,, \label{eq:NssMR}
\end{align}
\end{subequations}
where we have not written the $N_{,\p\s}$ component since we find that this is negligible. The rewriting of eq.~\eqref{eq:NppMR} is afforded by the assumption that $U_\p$ is the only contribution to the energy density of the Universe. We therefore see that the only way to obtain $\fnl$ larger than $\O(\ep^*)$ is from the $N_{,\s\s}$ term.

We note that the limit $\rho_\s \to 0$ does not invalidate the use of $r^{\rm C}$, because the function $f^{\rm C}$ can be defined to compensate for this apparently singular behaviour. Such details are automatically accounted for by simply writing all instances of $f^{\rm C}_{,\p}$ or $f^{\rm C}_{,\s}$ in terms of $r^{\rm C}$ and noting that $r^{\rm C}$ is directly related to the angle that the hypersurface makes in phase space which is a perfectly regular quantity.
A final point about eq.~\eqref{eq:NssMR} is that it does not lead to a divergence in $\fnl$ for $\gpd = \gpc$ because such divergent behaviour is regularized by the prefactor of $R^2$ in eq.~\eqref{eq:fnl} for $\fnl$.

From these general formulae we can make some more detailed inference. A plot of $\fnl$ against time will be a step function with a slow-roll suppressed value before $t^{\rm C}$ where it will jump in value. The final value depends on the value of $r^{\rm C}$. If we choose $r^{\rm C}=0$ then there is no modulation and so there is no step. As we increase the magnitude of $r^{\rm C}$ we obtain the following behaviour
\begin{itemize}
\item{$|r^{\rm C}|\ll1$}
\begin{equation}
\frac{N_{,\s\s}}{N_{,\s}^2} = 
\frac{3\gsc \gpd}{\gpd - \gpc} \bigg[
\F^{\rm C} +\frac{U_\s^{\rm B}}{{{U_\s'}^{\rm B}}^2} \bigg( 2{U_\s''}^{\rm B} - {U_\s''}^* \bigg)
\bigg] \frac{1}{r^{\rm C}} \,,
\end{equation}
\item{$r^{\rm C} = 1$}
\begin{equation}
\frac{N_{,\s\s}}{N_{,\s}^2} = 
\frac{3\gsc \gpd}{\gpd - \gpc} \bigg[
-1 +\frac{U_\s^{\rm B}}{{{U_\s'}^{\rm B}}^2} \bigg( 2{U_\s''}^{\rm B} - {U_\s''}^* \bigg)
\bigg]\,,
\end{equation}
\item{$|r^{\rm C}| \gg 1$}
\begin{equation}
\frac{N_{,\s\s}}{N_{,\s}^2} = 
\frac{3\gsc \gpd}{\gpd - \gpc} \bigg[
\F^{\rm C} + 1 - \frac{\gpc}{\gsc} 
\bigg]r^{\rm C} \,,
\end{equation}
\end{itemize}
where there is nothing special about $r^{\rm C}=1$ but we include this to illustrate the behaviour for intermediate values of $r^{\rm C}$. 

We see that $|\fnl| \gtrsim 1$ is possible for small values of $r^{\rm C}$ but this is only possible if $R$ is sufficiently large. To obtain $R \geq 1$ for quadratic potentials this requires $\s^* \leq r^{\rm C} \mpl/(6 \sqrt{N^{\rm A}})$ which represents an increasing degree of fine-tuning in $\s^*$ as $r^{\rm C} \to 0$.
Alternatively, for large values of $r^{\rm C}$ it is easy to obtain $R \geq 1$ without significant fine-tuning on $\s^*$ and, as expected, the sign of $\fnl$ is dependent on the geometry of the reheating hypersurface. Finally, for intermediate values of $r^{\rm C}$ we find small but non-negligible results such as the case with $r^{\rm C}=1$ and $U_\s$ as a quadratic potential in which case we obtain $\fnl = -5$.

\section{Conclusions}
\label{sec:conclusions}

This paper provides an analytic formalism for showing how perturbative reheating modifies the predictions of two-field slow-roll inflation. This is therefore an extension of the inflationary work carried out by Vernizzi and Wands~\cite{Vernizzi:2006ve}. Our calculations are made possible by replacing the oscillating fields with effective fluids. 

The hypersurfaces where the two fields begin to oscillate are arbitrary, as are the hypersurfaces where the two fields reheat. This allows us to include all of the effects of modulation throughout. In addition, we do not require that the fields reheat in any particular order. Despite this generality, the linear order contribution to $\zeta$ in eq.~\eqref{eq:dN_3} is reasonably simple. This linear order result is the principle result of this paper, since all higher order results follow by differentiation. We provide the ingredients needed to compute the second order results, which may be assembled to suit a given problem. Our explicit results are specific to the four phases illustrated in figure~\ref{fig:summary} which reduces to a large range of existing scenarios in different limits. The calculations that define our formalism are more general, however, and can easily be adapted to models that do not fit this four-phase picture, such as the inflating curvaton.

There is a clear physical motivation to consider modulation of the reheating hypersurfaces, since this allows us to fully account for the physical effects of modulated reheating or an inhomogeneous end of inflation. The main motivation to allow the fields to begin oscillations on an arbitrary hypersurface is simply that we do not currently have a robust prescription for what this hypersurface should be. Earlier work by Kawasaki et al.~\cite{Kawasaki:2012gg} demonstrated that this can be different from a surface of uniform density and they provided a simple prescription that works well for spectator models---where one field dominates the energy density. We then tested this prescription outside of the spectator regime and found that it produces spurious effects on the predictions for observables. Further work is therefore required to discern the analytic conditions that define the onset of oscillations. Since our formalism is agnostic about such analytic conditions, it provides a suitable test bed for studying different suggestions.

We have used our formalism to discuss predictions for a model where reheating is modulated by a fluid rather than a field. The conclusions show that a large value of $\fnl$ is easily produced if the hypersurface has the correct geometry. This is consistent with the results of Elliston et al.~\cite{Elliston:2013afa} where it was shown that the geometry of the reheating hypersurface is significantly constrained in the light of Planck bispectrum data. We therefore expect that this scenario may be similarly constrained.

Our work has important consequences for earlier numerical work. In particular, Leung et al.~\cite{Leung:2012ve} discussed the evolution of $\fnl$ as one varies the $\Gamma$ parameters of perturbative reheating. Like Meyers and Tarrant~\cite{Meyers:2013gua}, our work demonstrates that these numerical findings can be neatly explained by considering the relative redshifting of the two fluids involved. 

\section{Acknowledgements}
We are grateful to Ewan Tarrant and Joel Meyers for their constructive comments. JE and DJM are respectively supported by the Science and Technology Facilities Council grants ST/I000976/1 and ST/J001546/1. SO is supported by the Swiss National Science Foundation. 

\appendix

\section{Background dynamics of spectator models}
\label{sec:spec_back}

The spectator assumption ensures that the dominant $\p$ field evolves independently of the $\s$-component dynamics. Choosing $t^{\rm A}$ by the condition of Kawasaki et al.~\cite{Kawasaki:2012gg} as given in eq.~\eqref{eq:kawasaki}, we find $\p^{\rm A}\approx \sqrt{2} \mpl$. Integrating the background equation of motion for the $\p$ field then yields $\p^* \approx 2 \mpl \sqrt{N^{\rm A}}$. We therefore find $\rho_\p^{\rm A} = \frac{1}{2} m_\p^2 {\p^{\rm A}}^2 = \mpl^2 m_\p^2$. This allows us to find 
$\rho^{\rm B} \approx \rho_\p^{\rm B} = \mpl^2 m_\p^2 e^{-3 \gpc N^{\rm B}}$.
We can also apply eq.~\eqref{eq:kawasaki} to find the onset of $\s$ oscillations which provides the additional constraint $\rho_\p^{\rm B} = 3 \mpl^2 {U_\s'}^{\rm B} / (c^{\rm B} \s^{\rm B})$. We note that we find it necessary to modify eq.~\eqref{eq:kawasaki} in the vicinity of a hilltop at $\s_0$ such that $\s$-field oscillations begin when $|(\s-\s_0)' / (\s-\s_0)|$ equals unity rather than $|\s' / \s|$.
Combining these relations for $\rho_\p^{\rm B}$ we therefore find an important constraint on the parameters as
\begin{equation}
\label{eq:back_const}
e^{3 \gpc N^{\rm B}} = \frac{c^{\rm B} \s^{\rm B} m_\p^2}{3 {U_\s'}^{\rm B}}\,, \qquad
\frac{c^{\rm B}}{3} m_\p^2 \geq \frac{{U_\s'}^{\rm B}}{\s^{\rm B} } \,,
\end{equation}
where the inequality follows since $N^{\rm B} \geq 1$. A large hierarchy must exist between the two sides of this inequality if $N^{\rm B}$ is to take any value greater than one efold.
 
Let us now consider the dynamics of the $\s$ field during phases A and B. The background equations of motion are respectively given in eqs.~\eqref{eq:field_field} and eq.~\eqref{eq:field_fluid}. These can be combined, integrated and simplified using eq.~\eqref{eq:back_const} to yield
\begin{equation}
\label{eq:s_back}
\int_*^{\rm B} \! \frac{\d \s}{U_\s'} \approx -\frac{1}{2 m_\p^2} \bigg(
\ln (4 N^{\rm A}) + \frac{2 e^{3 \gpc N^{\rm B}}}{c^{\rm B} \gpc} \bigg)\,.
\end{equation}
In deriving this result we have neglected the two boundary terms on the right-hand-side that are evaluated at $t^{\rm A}$. Such terms only exist as an artifact of the imperfect matching between the two phases and we find that they are sufficiently small that they may be neglected. The two terms on the right hand side of eq.~\eqref{eq:s_back} are sourced respectively by the dynamics during phases A and B. Since we require $N^{\rm A} \approx 60$ to be consistent with observation, we find that the first term dominates for $N^{\rm B} \lesssim 1$ for reasonable equations of state and the second term dominates otherwise. For the hilltop model considered in \S\ref{sec:hilltopcurvaton} we specialise to the case where there is a significant mass hierarchy such that this second term dominates. We can then use eq.~\eqref{eq:back_const} to write
\begin{equation}
\label{eq:s_back2}
\int_*^{\rm B} \! \frac{\d \s}{U_\s'} \approx -
\frac{ \s^{\rm B} }{3 \gpc {U_\s'}^{\rm B}} \,.
\end{equation}
For a quadratic potential $U_\s = \frac{1}{2} m_\s^2 \s^2$ one then finds
\begin{equation}
\ln \bigg( \frac{\s^{\rm B}}{\s^*} \bigg) = -\frac{1}{3\gpc} \,,
\end{equation}
of for an inverted quadratic potential $U_\s = {\rm const.} - \frac{1}{2} m_\s^2 (\s - \s_0)^2$ one finds
\begin{equation}
\label{eq:sig_hilltop}
\ln \bigg( \frac{\s_0-\s^{\rm B}}{\s_0-\s^*} \bigg) = \frac{1}{3\gpc} \,.
\end{equation}

\bibliography{hilltop}

\end{document}